\documentstyle[12pt,a4,aps,epsfig,floats]{revtex}
\tighten

\def\ii{{\rm i}}	%imaginary unit
\def\de{{\rm\,d}}	%differential 
\def\e{{\rm e}}		%Neper's constant
\def\<{\langle}		%bra
\def\>{\rangle}		%ket
\def\tr{\mbox{Tr}}	%trace of
\def\^{\hat}		%hat

\newcommand{\bm}[1]{\mbox{\boldmath $#1$}}	%bold in formulae
\newcommand{\g}{\gamma}			
\newcommand{\sig}{\sigma}

\newcommand{\nn}{\nonumber}
\newcommand{\ovl}{\overline}

\newcommand{\Pslash}{\kern 0.2 em P\kern -0.56em \raisebox{0.3ex}{/}}
\newcommand{\Sslash}{\kern 0.2 em S\kern -0.56em \raisebox{0.3ex}{/}}
\newcommand{\pslash}{\kern 0.2 em p\kern -0.4em /}
\newcommand{\kslash}{\kern 0.2 em k\kern -0.45em /}
\newcommand{\qslash}{\kern 0.2 em q\kern -0.4em /}

\newcommand{\xbj}{x_{\scriptscriptstyle B}}     %x Bjorken            

\begin{document}

%%%%%%%%%%%%%%%%%%%%%%%%%%%%%%%%%%%%%%%%%
\title{
\begin{flushright}
\begin{minipage}{4 cm}
\small
VUTH 00-8\\
FNT/T-00/04\\
WU-B 00-08\\
\end{minipage}
\end{flushright}
Semi-inclusive structure functions in the spectator model}

\author{A.~Bacchetta$^1$, S.~Boffi$^2$, R.~Jakob$^3$}

\address{
$^1$Division of Physics and Astronomy, Faculty of Science, Vrije Universiteit\\
De Boelelaan 1081, 1081 HV Amsterdam, the Netherlands\\[2mm]
$^2$Dipartimento di Fisica Nucleare e Teorica,
Universit\`{a} di Pavia and INFN, Sezione di Pavia\\ 
Via Bassi 6, 27100 Pavia, Italy\\[2mm]
$^3$Fachbereich Physik, Universit\"{a}t Wuppertal\\
Gauss-Stra{\ss}e 20, 42097 Wuppertal, Germany}

\date{draft of \today}
\maketitle

%%%%%%%%%%%%%%%%%%%%%%%%%%%%%%%%%%%%%%%%%%%%%%5
\begin{abstract}
We establish the relationship between distribution and fragmentation functions
and the structure functions appearing in the cross section 
of polarized 1-particle inclusive deep-inelastic scattering.
We present spectator model evaluations of these structure functions
focusing on the case of an outgoing 
spin-$\frac{1}{2}$ baryon. Distribution functions obtained in 
the spectator model are known to fairly agree at low energy scales 
with global parameterizations extracted, for instance, from totally
inclusive DIS data.
Therefore, we expect it to give good hints on the functional dependence of the
structure functions on the scaling variables $\xbj$, $z$ and on the transverse
momentum of the observed outgoing hadron, ${P}_{h\perp}$. Presently, this
dependence is not very well known, but experiments are planned in  the near
future. 
\end{abstract}

%%%%%%%%%%%%%%%%%%%%%%%%%%%%%%%%%%%%%%%%%
\section{Introduction}
Totally inclusive deep-inelastic scattering (DIS) in the past years provided 
us with
rather precise knowledge of the distribution functions of the proton and the
neutron, helping us to understand their inner structure and raising questions
yet 
unanswered. Semi-inclusive DIS displays even richer characteristics. Detecting 
at least one of the hadrons produced in the high-energy scattering 
process and measuring its momentum, one is then sensitive not only to the 
distribution of partons inside the target hadron, but also to the mechanism of 
hadronization, through which a quark gives rise to a jet of new 
hadrons. We are then
able to measure not only the distribution functions, but also the so-called 
fragmentation functions. Neither the distribution nor the fragmentation 
functions can be calculated from first principles within perturbative QCD, 
because they belong to the 
non-perturbative realm of bound states. Therefore, models are required. 

If only the dominant light-cone 
component of the momentum of the outgoing 
hadron is measured, its transverse components being integrated over, then the 
structure functions appearing in cross-sections will be products of a 
distribution and a fragmentation function. The already established 
knowledge of the distribution functions enables one to extract the shape 
of the fragmentation functions~(cf.~\cite{EMC}) and to
compare it with other results coming from different experiments, such as 
electron-positron annihilation.

On the other hand, if we manage to measure the transverse momentum of the 
outgoing hadron we have the opportunity to study some new interesting 
distribution and fragmentation functions. In particular, already to leading
order in an expansion in powers of $\frac{1}{Q}$ we have access 
to chiral odd and time-reversal odd functions, as well as 
functions related to the transverse momentum carried by quarks relative to
their parent hadrons momentum~\cite{mul}. These functions
are presently considered to be very interesting and their experimental
measurement is in progress (HERMES, COMPASS, RHIC). The only major
inconvenience of dealing  with cross sections differential in the transverse
momentum of the outgoing hadron is that the structure functions are no more
simple products of a distribution and a fragmentation function, but rather
convolutions of  those. 

In this context model evaluations of the structure functions can be 
very useful. The spectator model proved to be in qualitative agreement 
with the known (transverse momentum integrated)
distribution and fragmentation functions evolved at low energies \cite{rod}. 
Therefore, we expect it to give reasonable estimates for the convolution
integrals in semi-inclusive DIS, provided that the inclusion of transverse 
momentum of partons does not spoil factorization properties, as it is
usually assumed \cite{col}.

We are aware that our results cannot be considered as precise
predictions of experimental quantities, because of the limitations and the
simplicity of the model.

The model incorporates only valence 
quark distribution and fragmentation, neglecting the presence of sea-quarks,
gluons and evolution. Therefore, it is supposed to reproduce the shape of the
valence-quark distribution and fragmentation at a low energy scale, which is
not known a priori.  To give an estimate of this scale, one can compare the 
total momentum carried by quarks as given by the
spectator model with the same quantity as given by parameterizations of
the distribution functions at a known energy scale. Such a comparison suggests
that the spectator model is valid at an energy scale of about 0.2-0.3 GeV. 

In principle, the results we obtain need to be evolved to higher energy
scales by means of evolution equations for a final comparison with experiment.
Evolution equations for transverse momentum dependent functions are not yet
known. In this article, therefore, we refrain from taking into account
radiative  corrections.
Nevertheless, we are confident to describe correctly the main features and
properties of the structure functions at low (intrinsic) transverse momentum,
since perturbative corrections are expected to affect mainly the high
transverse momentum tails of these functions.

The spectator model is a semi-phenomenological model, which relies mainly on
the idea of describing the hadron as an ensemble of a free parton (struck 
in the scattering process) and a fictitious, unphysical particle, the 
spectator, with the right quantum numbers. 

The model, at least in its present version, cannot describe time-reversal 
odd functions, since it does not incorporate final state interactions. On 
the other hand, the advantages of the model resides in the fact that
it is simple, it is covariant and it gives a clear estimate of the 
distribution of partonic transverse momentum. Last but non least, the model 
can be treated in wide parts analytically. Some numerical integrations are 
required only when the cross section is kept differential also in the 
transverse momentum of the produced hadron.

In Sec.~\ref{s:sidis}  we briefly review the general formalism utilized
to treat semi-inclusive DIS, with emphasis on the structure functions 
calculable using the spectator model. In Sec.~\ref{s:main} we 
present the basic properties of the model and  we 
give the results of
our analysis, highlighting what are the broad features that could possibly be 
observed in experiments.

%%%%%%%%%%%%%%%%%%%%%%%%%%%%%%%%%%%%%%%%%%%%%%%%%%%%%%%%%5
\section{Semi-inclusive DIS}
\label{s:sidis}

The cross section for a semi-inclusive DIS event can be written
in terms of the contraction between a lepton and a hadron tensor.  For instance
in the target rest frame we can make use of the formula \cite{lev}
\begin{equation}
\frac{\de \sig}{\de E' \de \Omega \de^3 P_h}=
		\frac{\alpha^2}{Q^4}\; \frac{E'}{E}\;
			    L_{\mu \nu}W^{\mu \nu}
\end{equation}
where $P_h$ is the momentum of the outgoing hadron,$Q^2=-q^2 \geq 0$ is the 
absolute value of the virtuality of the exchanged photon,  $\alpha=e^2/4\pi$ 
is the fine structure constant,
$E$ and $E'$ are the energy of the incident
lepton before and after the collision, respectively. $L_{\mu \nu}$ is the 
lepton tensor and $W^{\mu \nu}$ is the hadronic tensor.  

Following a purely phenomenological approach, the hadronic tensor can be 
parameterized using  scalar structure functions. 
A priori, the maximum number of independent structure functions in 
an arbitrary DIS process is 16. Since we will be
interested only in electromagnetic scattering, this number is reduced to 9 by 
the gauge invariance condition,
$q_{\mu}W^{\mu \nu}= q_{\nu}W^{\mu \nu}=0$. A convenient set of functions is
formed by the  spherical basis structure functions (see, e.g., \cite{bof}).

In the spherical basis, constraints coming from angular momentum conservation 
take a simpler form.
From now on, we want to consider only leading terms in an expansion over 
$\frac{1}{Q}$. In the case of polarized semi-inclusive scattering, helicity
conservation considerations allow us to say that five of the structure 
functions vanish at leading order, leaving only four
non-zero structure functions. The complete form of the hadronic tensor at
leading order is then
\begin{equation}
W^{\mu\nu} = -g^{\mu\nu}_{\perp}\frac{W_{T}}{2}
         +\ii\, \varepsilon^{\mu\nu}_{\perp}\frac{W'_{TT}}{2} 
	+\left( 2 \^P_{h \perp}^{\mu} \^P_{h \perp}^{\nu} +
			g^{\mu\nu}_{\perp}\right) \frac{W_{TT}}{2}
	+\^P_{h \perp}^{\{\mu}\varepsilon^{\nu\}\rho}_{\perp}
			\^P_{h \perp\;\rho}\frac{\ovl{W}_{TT}}{2} ,
\label{e:strutt2}
\end{equation}
where the curly brackets indicate symmetrization of the indices. For the
definitions of the tensor structures appearing in the formula we need to
define a normalized time-like and a normalized space-like vector
\begin{equation} 
\^t^\mu=\frac{2\xbj}{Q}\left(P^\mu-q^\mu\frac{P\cdot q}{q^2}\right) \;,
\qquad
\^q^\mu=\frac{q^\mu}{Q} \;,
\end{equation}
by means of which we can define the structures
\begin{eqnarray} 
g_\perp^{\mu\nu}&=&g^{\mu\nu}+\^q^\mu\^q^\nu-\^t^\mu\^t^\nu \;, \\
\varepsilon^{\mu\nu}_{\perp}&=&
		\varepsilon^{\mu\nu\alpha\beta}\^q_{\alpha}\^{t}_{\beta}\;, \\
\^P_{h \perp}^{\mu}&=& \frac{g_\perp^{\mu\rho}P_{h\;\rho}}
			   {|g_\perp^{\mu\rho}P_{h\;\rho}|}\;.
\end{eqnarray}

\begin{figure}
        \centering
        \epsfig{figure=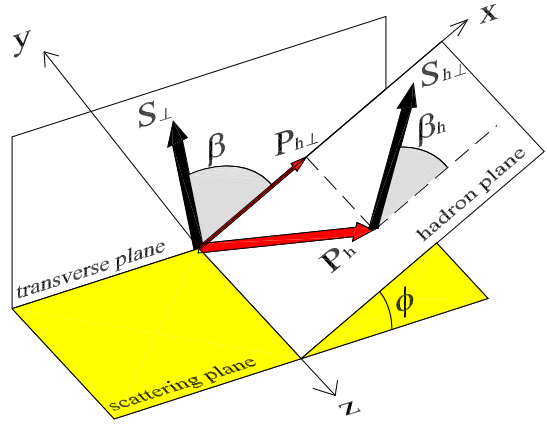,height=6cm}
        \caption{Sketch of the angles used in the description of the hadronic
		tensor}
        \label{f:beta}
        \end{figure}

By calculating the contraction between leptonic and hadronic tensor we can
eventually write the following formula for the cross section \cite{bof} in the
target rest-frame:
\begin{equation}
\frac{\de \sig}{\de E' \de \Omega \de^3 P_h}= 
\sigma_M \;{Q^2\over 2 \vert \bm{q}\vert^2}\,{1\over\epsilon}
\left\{W_T 
+ \epsilon\,\left(W_{TT}\,\cos{2\phi} + \overline{W}_{TT}\,\sin{2\phi}\right)
+ \lambda_e\, \sqrt{1-\epsilon^2}\; W'_{TT} 
\right\},
\end{equation} 
where $\lambda_e$ is the helicity of the electron, $\phi$ is the angle between
the scattering plane and the outgoing hadron's momentum (see Fig.~\ref{f:beta})
and where
\begin{eqnarray} 
\sigma_M &=& \frac{4\alpha^2 {E'}^{2}}{Q^4} 
		\,\cos^2{\left(\frac{\theta}{2}\right)}, \nn \\
\epsilon^{-1} &=& 1+ 2\, {\vert \bm{q}\vert^2 \over Q^2} 
		\,\tan^2{\left(\frac{\theta}{2}\right)},
\end{eqnarray}
$\theta$ being the scattering angle of the electron.

\begin{figure}
        \centering
        \epsfig{figure=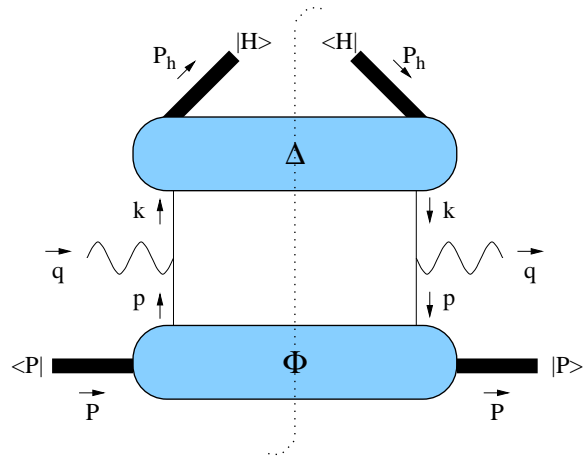,height=6cm}
        \caption{Diagrammatic representation of semi-inclusive DIS}
        \label{f:scat}
        \end{figure}

The dominant contribution to the structure functions can be calculated from
the cut-diagram shown in Fig.~\ref{f:scat}, representing the
hadronic part of a DIS event.

According to  the usual factorization assumption, the diagram is divided into a
hard partonic scattering amplitude and two soft parts, $\Phi$ and 
$\Delta$ in Fig.~\ref{f:scat}, called correlation functions. 
The dominant momentum 
direction in the upper (lower) part is given by the direction of the outgoing 
(target) hadron momentum, $P_h$ ($P$). Quark momenta are almost collinear
to their parent hadrons allowing for small transverse components.
	Using Feynman rules we can give an explicit formula describing the 
diagram of Fig.~\ref{f:scat}, which represents the hadronic tensor:
\begin{equation}
2MW^{\mu \nu}= \left.\int \de k^+ \de^2{\bm k}_T \de p^-
        \de^2{\bm p}_T \;\delta^{(2)}({\bm p}_T+{\bm q}_T-{\bm k}_T)\sum_q
e_q^2 \; 			\tr [ \Phi_q\, \g^{\mu}\, \Delta_q \,  
        \g^{\nu}]\right|_{\stackrel{{\scriptstyle 
			p^+={\xbj} P^+}}
        {k^-=P_h^-/z}},
\label{e:w1}
\end{equation}
where $p$ and $k$ are the momenta of the quarks respectively before and after
absorbing the photon, and the index $q$ denotes quark flavor.

In this formula we use light-cone components of vectors in an infinite
momentum  frame of 
reference where the ``$+$'' direction is given by the momentum of the target 
hadron, $P$, the ``$-$'' direction is given by the momentum of the outgoing 
hadron, $P_h$, and the photon momentum is purely spatial. 
In this frame of reference the incident photon has a transverse
component, ${\bm q}_T$. In alternative, one can work with 
frames of reference where the photon does not have transverse components,
which are particularly convenient from the experimental point of view. In  such
frames the outgoing hadron's momentum acquires a transverse  component, which
we will denote as ${\bm P}_{h\perp}$.  It can be shown that the relation
between these transverse components is  \cite{mul}:
\begin{equation}
{\bm q}_T=-\frac{{\bm P}_{h\perp}}{z}.
\label{e:qTpperp}
\end{equation}

In Eq.~(\ref{e:w1}) the already mentioned correlation functions appear. 
They are second-rank Dirac tensors defined as:
\begin{eqnarray}
\Phi_{q(mn)}(p,P,S)
&=& \int \frac{\de^{4}x}{(2\pi)^{4}}
                \e^{- \ii px}
        \<P,S|\bar{\psi}^q_{(n)}(x)\psi^q_{(m)}(0)|P,S\>,\\
\Delta_{q(mn)}(k,P_h,S_h)&=&\int
        \frac{\de^{4}x}{(2\pi)^{4}}\e^{+\ii kx}
        \<0|\psi^q_{(m)}(x)|P_h,S_h\>\<P_h,S_h|
             \bar{\psi}^q_{(n)}(0)|0\>.    
\label{e:corr}
\end{eqnarray}

If we decompose each correlation function on a basis of 16 Dirac structures, 
$\Gamma_i$, we get the following result:
\begin{equation}
2MW^{\mu \nu}
	= \sum_{i,j}\frac{\tr[\Gamma^i \g^{\mu} \Gamma^j \g^{\nu}]}{4}\;
           2z \int \de^2{\bm k}_T \de^2{\bm p}_T \,\delta^{(2)}({\bm p}_T+{\bm
q}_T-{\bm k}_T)\,\sum_q e_q^2 \;                 \Phi^{[\Gamma_i]}_q 
\Delta^{[\Gamma_j]}_q. \label{e:w2}
\end{equation}
where we defined
\begin{eqnarray}
\Phi^{[\Gamma^i]}_q(\xbj ,{\bm p}_T)& =& \makebox[6cm][l]
	{$\left. \frac{1}{2}  \int \de p^-\;\tr [\Phi_q
		\Gamma^i] \right|_{p^+={\xbj} P^+}$}
	\mbox{(distribution functions)}, 
\label{e:dist}\\
\Delta^{[\Gamma^i]}_q(\frac{1}{z} ,{\bm k}_T)& =&\makebox[6cm][l]
	{$ \left. \frac{1}{4z} \int \de k^+ \;
                    \tr[\Delta_q \Gamma^i] \right|_{k^-=\frac{P_h^-}{z}}$}
	\mbox{(fragmentation functions)}. 
\label{e:frag}
\end{eqnarray}

As a last step, the distribution and fragmentation functions are usually 
divided in several components,
named $f_1$, $g_{1L}$, $g_{1T}$, \ldots for what concerns the distribution 
functions, or $D_1$, $G_{1L}$, $G_{1T}$, \ldots for what concerns the 
fragmentation functions. We will not pursue the definition of all the possible
functions, for which we simply refer to \cite{mul}. Note that each function
should always carry its flavor index, $q$, although in the rest of this section
we are going to skip it.

Using Eq.~(\ref{e:w2}) and comparing it with Eq.~(\ref{e:strutt2}) 
we can establish relations connecting structure
 functions to distribution and fragmentation functions.
In the following equations we will use the frame of reference and the angles
specified in Fig.~\ref{f:beta}.
 We chose to fix the $x$ axis in the direction of
${\bm P}_{h\perp}$ to simplify the formulae. Note that we do not lose 
generality
because in the hadronic tensor there is no dependence on the angle $\phi$ due
to cylindrical symmetry around the $z$ axis.

To be more concise we will denote the convolution
integral appearing in the hadronic tensor  and the summation over quark
flavors with the following notation:
\begin{equation} 
	{\cal I}\left\{\,\ldots\,\right\}=
 		 \frac{2z}{M} \int \de k_x \de k_y  \de p_x \de p_y \;
	\delta\left(p_x- \frac{|P_{h\perp}|}{z}-k_x\right) \delta(p_y-k_y)
              \,\sum_q e_q^2  \,\ldots\,,	\label{e:convo2}
\end{equation}
where the specific form of the $\delta$-functions is due to
relation~(\ref{e:qTpperp}) and to the particular choice of our $x$ axis.

We are going to divide the structure functions in various 
contributions arising from particular polarization conditions.
Therefore, we are going to label each contribution with two indices,
 the first one referring to the 
polarization of the target ($U$ for unpolarized, $L$ for longitudinally 
polarized, $T$ for transversely polarized) and the second one referring to the 
polarization of the outgoing hadron. We will not take into account the cases 
when both hadrons are polarized. 

The resulting form of the structure functions is 
\begin{eqnarray} 
W_{T}(\xbj,z,P_{h\perp}) &=& 
	W_T^{\,\mbox{\tiny[UU]}}(\xbj,z,P_{h\perp})+	
      |S_{h \perp}| \sin{\beta_h}\;W_T^{\,\mbox{\tiny[UT]}}(\xbj,z,P_{h\perp}),
	\label{e:wt}\\ \nn \\
&&\hspace{-1cm} W_T^{\,\mbox{\tiny[UU]}}(\xbj,z,P_{h\perp})\,=\, 
	{\cal I}\left\{\,f_1 D_1\,\right\},	\label{e:wtuu}	\\ 
&&\hspace{-1cm} W_T^{\,\mbox{\tiny[UT]}}(\xbj,z,P_{h\perp})\,=\,
	 {\cal I}\left\{\,\frac{k_x}{M_h}f_1 D_{1T}^{\perp}\,\right\},
						\label{e:wtut}	\\
		\nn \\ \nn \\
W_{TT} (\xbj,z,P_{h\perp}) &=&
   |S_{\perp}| \sin{\beta}\; W_{TT}^{\,\mbox{\tiny[TU]}}(\xbj,z,P_{h\perp}), \\
		\nn \\
&&\hspace{-1cm} W_{TT}^{\,\mbox{\tiny[TU]}}(\xbj,z,P_{h\perp})\,=\,
	{\cal I}\left\{\,\frac{k_x}{M_h}
	\,h_{1T} H_1^{\perp}\,+\frac{k_y p_y p_x + k_xp_yp_y}{2 M^2M_h}
	\,h_{1T}^{\perp}H_1^{\perp}\,\right\},   \label{e:wtttu}	\\
		\nn \\ \nn \\
{W'}_{TT}(\xbj,z,P_{h\perp})  &=& 
	\lambda\; {W'}_{TT}^{\,\mbox{\tiny[LU]}}(\xbj,z,P_{h\perp})+
    |S_{\perp}|\cos{\beta}\;{W'}_{TT}^{\,\mbox{\tiny[TU]}}(\xbj,z,P_{h\perp}) 
	\nn\\&&\mbox{}+
   \lambda_h\; {W'}_{TT}^{\,\mbox{\tiny[UL]}}(\xbj,z,P_{h\perp})+
|S_{h\perp}| \cos{\beta_h}\; {W'}_{TT}^{\,\mbox{\tiny[UT]}}(\xbj,z,P_{h\perp}),
	\phantom{\frac{k_x}{M_h}}    \label{e:wtt'}\\	\nn \\
&&\hspace{-1cm} {W'}_{TT}^{\,\mbox{\tiny[LU]}}(\xbj,z,P_{h\perp})\,=\,
	{\cal I} \left\{\, g_{1L} D_1\,\right\},  \label{e:wtt'lu}\\
&&\hspace{-1cm} {W'}_{TT}^{\,\mbox{\tiny[TU]}}(\xbj,z,P_{h\perp})\,=\,
	{\cal I} \left\{\,\frac{p_x}{M} \,g_{1T} D_1\,\right\}, 
						\label{e:wtt'tu}\\
&&\hspace{-1cm} {W'}_{TT}^{\,\mbox{\tiny[UL]}}(\xbj,z,P_{h\perp})\,=\,
	{\cal I}\left\{\, f_1 G_{1L}\,\right\},  \label{e:wtt'ul}\\
&&\hspace{-1cm} {W'}_{TT}^{\,\mbox{\tiny[UT]}}(\xbj,z,P_{h\perp})\,=\,	
	{\cal I}\left\{\,f_1 \frac{k_x}{M_h}\,G_{1T}\,\right\}, 
						\label{e:wtt'ut}\\
		\nn \\ \nn \\
\ovl{W}_{TT}(\xbj,z,P_{h\perp})  &=&  
	\lambda\; \ovl{W}_{TT}^{\,\mbox{\tiny[LU]}}(\xbj,z,P_{h\perp})
	-|S_{\perp}| \cos{\beta} \; 
		\ovl{W}_{TT}^{\,\mbox{\tiny[TU]}}(\xbj,z,P_{h\perp}),
		\\ \nn \\
&&\hspace{-1cm} \ovl{W}_{TT}^{\,\mbox{\tiny[LU]}}(\xbj,z,P_{h\perp})\,=\,
	{\cal I}\left\{\,\frac{k_xp_x-k_yp_y}{M M_h}
	 h_{1L}^{\perp} H_1^{\perp}\,\right\},   \label{e:wttbarlu}\\
&&\hspace{-1cm} \ovl{W}_{TT}^{\,\mbox{\tiny[TU]}}(\xbj,z,P_{h\perp})\,=\,
	{\cal I}\left\{\,\frac{k_x}{M_h}\,h_{1T} H_1^{\perp}\,+
		\frac{k_x p_x p_x - k_y p_y p_x}{M^2 M_h}
		\,h_{1T}^{\perp}H_1^{\perp}\,\right\}.	\label{e:wttbartu}
\end{eqnarray}

Here, $\lambda$ ($\lambda_h$) and $S_{\perp}$ ($S_{h \perp}$) denote helicity
and transverse spin of the target (outgoing) spin-$\frac{1}{2}$ hadron.

The distribution functions (small letters) are understood to be functions 
of the variables $\xbj$ and ${\bm p}_T^2$, while the fragmentation functions 
(capital letters) are understood to be functions of 
$\frac{1}{z}$ and ${\bm k}_T^2$.
		
The functions $D_{1T}^{\perp}$ and $H_1^{\perp}$ are time-reversal odd and they
cannot be studied in the framework of our model. For this reason we will only 
be able to calculate the first term of the structure function $W_T$ and 
the full structure function $W'_{TT}$.

Due to Eq.~(\ref{e:qTpperp}), integrating over the outgoing hadron transverse 
momentum, ${\bm P}_{h\perp}$, corresponds to 
integrating over $z^2 {\bm q}_T$.
Performing this integration leads to a  
deconvolution of the right-hand side of Eq.~(\ref{e:w2}). Consequently, the 
integrals over ${\bm p}_T$ and ${\bm k}_T$ can be performed separately for the
distribution and fragmentation functions.
Therefore, the structure functions integrated over ${\bm P}_{h\perp}$  
reduce to:
\begin{eqnarray}
W_{T\phantom{T}}(\xbj,z)  &=&\frac{1}{M}\;2z
	\,f_1(\xbj) D_1({\textstyle \frac{1}{z}}),   
		\label{e:wt2}\\
\nn \\ 
W'_{TT}(\xbj,z)  &=& \frac{\lambda}{M}\;2z
	\, g_{1}(\xbj) D_1({\textstyle \frac{1}{z}})
   	\mbox{}+\frac{\lambda_h}{M}\;2z
	\, f_1(\xbj) G_{1}({\textstyle \frac{1}{z}}),
			             	\label{e:wtt'2}
\end{eqnarray} 
where the new ${\bm p}_T$ and 
${\bm k}_T$-independent distribution and fragmentation functions are defined 
as:
\renewcommand{\arraystretch}{1.5}
\begin{equation} 
\begin{array}{rclcrcl}
f_1(\xbj) &\equiv& 
{\displaystyle  
{\displaystyle \int}} \de^2 {\bm p}_T\; f_1(\xbj, {\bm p}_T^2),  &\qquad&
g_1(\xbj) &\equiv&  
{\displaystyle \int} \de^2 {\bm p}_T\; g_{1L}(\xbj, {\bm p}_T^2),  \\
D_1({\textstyle \frac{1}{z}}) &\equiv& z^2  
{\displaystyle \int} \de^2 {\bm k}_T\; 
                   D_1({\textstyle \frac{1}{z}}, {\bm k}_T^2),   &\qquad&
G_1({\textstyle \frac{1}{z}}) 
	&\equiv& z^2  
{\displaystyle \int} \de^2 {\bm k}_T\; 
			G_{1L}({\textstyle \frac{1}{z}}, {\bm k}_T^2).
\end{array}
\end{equation}

%%%%%%%%%%%%%%%%%%%%%%%%%%%%%%%%%%%%%%%%%%%%%%%%%%%%%%%%%%%%%%%%%%%%%%%%%%%%%%
\section{Calculation of structure functions}
\label{s:main}

\subsection{The spectator model}
\label{s:diquark}

For a numerical evaluation of the structure functions we will employ the
distribution and fragmentation functions as estimated with a spectator model
\cite{rod}. 
The basic assumption of the spectator model is that the target hadron can be
divided 
into  a quark and an effective spectator state with the required quantum 
numbers, which is treated to a first approximation as being on-shell with a
definite mass. In the case of a baryon target, this second particle is a 
diquark.

The same idea applies to the hadronization process: the quark fragments into a
jet, from which one hadron is eventually detected; the remnants of the jet are
treated effectively as an on-shell spectator state. If the detected 
hadron is a
baryon, the second particle is an anti-diquark.

The vertex coupling the baryon to quark and diquark includes a form factor
preventing the quark from being far off-shell. The large $p^2$-behavior of the
form factor is controlled by a parameter $\Lambda$.

We quote the analytic form of the distribution and fragmentation function we
are going to use for numerical evaluation of the structure functions as
obtained in \cite{rod}.   The diquark's spin in the simplest approach can be
either $0$ (scalar diquark with mass $M_s$) or $1$ (axial vector diquark with
mass $M_a$).  
For both cases the functions can be cast in the same analytic form where only
some parameters take different values. Therefore, we label the functions with
an additional index $i \in \{s,a\}$ to distinguish between the two cases. The
functions we consider are:
\begin{eqnarray}
f_1^i({\xbj},{\bm p}_T^2)&=&\frac{n_i^2 (1-{\xbj})^{3}}{16 \pi^3}\;
      \frac{(m + {\xbj}M)^2 + {\bm p}_T^2 }{({\bm p}_T^2 + l_i^2(x) )^{4}},
				\label{e:f1}\\ \nn \\
g_{1L}^i({\xbj},{\bm p}_T^2)&=&a_i\,\frac{n_i^2 (1-{\xbj})^{3}}{16 \pi^3}\;
   \frac{(m + {\xbj}M)^2 - {\bm p}_T^2 }{({\bm p}_T^2 + l_i^2(x) )^{4}},
			\label{e:g1l}	\\ \nn \\
g_{1T}^i({\xbj},{\bm p}_T^2)&=&a_i\,\frac{n_i^2 (1-{\xbj})^{3}}{16 \pi^3}\;
   \frac{2 M(m + {\xbj}M)}{({\bm p}_T^2 + l_i^2(x) )^{4}}, \label{e:g1t}
\\ \nn \\
D_1^i({\textstyle \frac{1}{z}},{\bm k}_T^2)&=&
	\frac{N_i^2 (1-z)^{3}}{16 \pi^3 z^4}\;
	\frac{(m + \frac{1}{z}M_h)^2 + {\bm k}_T^2  }
			{({\bm k}_T^2 + L_i^2(z))^{4}},
				\label{e:d1}\\ \nn \\
G_{1L}^i({\textstyle \frac{1}{z}},{\bm k}_T^2)&=&
	a_i\,\frac{N_i^2 (1-z)^{3}}{16 \pi^3 z^4}\;
	\frac{(m + \frac{1}{z}M_h)^2 - {\bm k}_T^2 }
			{({\bm k}_T^2 +L_i^2(z) )^{4}},	\label{e:G1l}
				\\ \nn \\
G_{1T}^i({\textstyle \frac{1}{z}},{\bm k}_T^2)&=&
	a_i\,\frac{N_i^2 (1-z)^{3}}{16 \pi^3 z^4}\;
	\frac{2 M_h(m + \frac{1}{z}M_h)}{({\bm k}_T^2 + L_i^2(z) )^{4}},
						\label{e:G1t}
\end{eqnarray}
with the spin factors $a_s=1$ and $a_a= -\frac{1}{3}$, and where we made use
of the newly defined functions: 
\begin{eqnarray} 
l_i^2(\xbj)& =& \Lambda^2(1-\xbj) + \xbj M_{i}^2 -\xbj (1-\xbj) M^2, \\
L_i^2(z)& =&\Lambda^2(1-{\frac{1}{z}}) 
	+ {\frac{1}{z}} M_{i}^2
	- {\frac{1}{z}} (1-{\frac{1}{z}}) M_h^2.
\end{eqnarray} 

The values of the parameters of the model have been determined to be
\begin{eqnarray}
\Lambda&=& 0.5 \mbox{ GeV}, \nn \\
M_{s}&=& 0.6 \mbox{ GeV},\qquad M_{a}= 0.8 \mbox{ GeV}, \\
m &=& 0.36 \mbox{ Gev}. \nn
\end{eqnarray}
The functions depend only weakly on the chosen value of the quark mass
$m$.
The normalization factors $n_i$ and $N_i$ are fixed by the conditions
\begin{eqnarray}  
\int \de \xbj \de^2 {\bm p}_T\; f_1^i(\xbj,{\bm p}^2_T)  &=& 1, \nn \\
\int \de z \de^2 {\bm k}_T\; z\, 
	\left(z^2 D_1^i({\textstyle \frac{1}{z}},{\bm k}^2_T)\right)  &=& 1.
\end{eqnarray} 
Note that for the fragmentation function the normalization condition is put 
on the first moment.
 
It is noteworthy to observe that the model intrinsically describes 
the dependence
of distribution and fragmentation functions on 
partonic transverse momentum. Moreover, this dependence is
significantly different from the frequently used Gaussian dependence.

We gave the distribution and fragmentation functions for scalar and
axial-vector diquarks. We must now address the problem of defining the hadron
state in terms of the quark-diquark content. 
From a group
 theory analysis, for a proton, a neutron and a $\Lambda$ particle the
results are \cite{kro}:
\begin{eqnarray} 
|p\> &=& 
	  {\textstyle {\frac{1}{\sqrt2}}} |u, S\> 
	+ {\textstyle {\frac{1}{\sqrt6}}} |u, A\> 
	- {\textstyle {\frac{1}{\sqrt3}}} |d, A\> ,
\label{e:protsa}  \nn \\
|n\> &=&
	  {\textstyle {\frac{1}{\sqrt2}}} |d, S\> 
	- {\textstyle {\frac{1}{\sqrt6}}} |d, A\> 
	+ {\textstyle {\frac{1}{\sqrt3}}} |u, A\> ,
\label{e:neutsa} \\
|\Lambda\> &=& 
	  {\textstyle {\frac{1}{\sqrt{12}}}}|u, S\> 
	- {\textstyle {\frac{1}{\sqrt{12}}}}|d, S\>  
	- {\textstyle {\frac{2}{\sqrt{12}}}}|s, S\>  
	+ {\textstyle {\frac{1}{\sqrt4}}} |u, A\> 
	- {\textstyle {\frac{1}{\sqrt4}}} |d, A\> .
\label{e:lamsa} \nn
\end{eqnarray} 

Using these results, the probability of finding an up, down or strange quark in
one of these hadrons is related to the probability of finding a scalar or
axial-vector diquark in the following way:
\begin{equation}
\begin{array}{rclcrcl} 
f_1^{p\rightarrow u} &=& \frac{3}{2}f_1^{s} + \frac{1}{2}f_1^{a}, &\qquad&  
f_1^{p\rightarrow d} &=& f_1^{a},				 \\
f_1^{n\rightarrow u} &=& f_1^{a},	&\qquad&			
f_1^{n\rightarrow d} &=& \frac{3}{2}f_1^{s} + \frac{1}{2}f_1^{a},   
					\\
f_1^{\Lambda\rightarrow u} &=& \frac{1}{4}f_1^{s} + \frac{3}{4}f_1^{a},   
					&\qquad&
f_1^{\Lambda\rightarrow d} &=& \frac{1}{4}f_1^{s} + \frac{3}{4}f_1^{a},	
					\qquad
f_1^{\Lambda\rightarrow s} = f_1^{a}.		
\end{array}
\label{e:nd}
\end{equation}
Overall normalizations ensure the correct number sum rules for valence quarks
in the baryons. Analogous formulae hold for all other distribution
functions and for the fragmentation functions as well. 

Once we have computed the distribution and fragmentation functions for a
scalar and axial-vector diquark, we can eventually calculate the argument of
the convolutions occurring in the structure functions for a given
process. At this stage, the (charge squared weighted) sum over quark flavors
is rewritten in a weighted sum over the different diquark species:
\begin{equation}
\sum_q e_q^2 f_1^q D_1^q = \sum_{i,j=s,a} c_{ij}\; f_1^i D_1^j,
\end{equation}
where the coefficients depend on the type of hadrons involved.

For instance, for the following processes:
\begin{itemize}
\item{$ e\,p\,\rightarrow e'\,\Lambda\,X$
\begin{equation} 
\frac{4}{9} f_1^{p\rightarrow u} D_1^{u \rightarrow \Lambda} + 
\frac{1}{9} f_1^{p\rightarrow d} D_1^{d \rightarrow \Lambda} =
	\frac{1}{6}f_1^s D_1^s +  \frac{1}{2}f_1^s D_1^a +
	\frac{1}{12}f_1^a D_1^s +  \frac{1}{4}f_1^a D_1^a; 
\label{e:protlam}
\end{equation}}
\item{$ e\,p\,\rightarrow e'\,p'\,X$
\begin{equation} 
\frac{4}{9} f_1^{p\rightarrow u} D_1^{u \rightarrow p} + 
\frac{1}{9} f_1^{p\rightarrow d} D_1^{d \rightarrow p} =
	f_1^s D_1^s +  \frac{1}{3}f_1^s D_1^a +
	\frac{1}{3}f_1^a D_1^s + \frac{2}{9}f_1^a D_1^a; 
\end{equation}}
\item{$ e\,n\,\rightarrow e'\,\Lambda\,X$
\begin{equation} 
\frac{4}{9} f_1^{n\rightarrow u} D_1^{u \rightarrow \Lambda} + 
\frac{1}{9} f_1^{n\rightarrow d} D_1^{d \rightarrow \Lambda} =
	\frac{1}{24}f_1^s D_1^s +  \frac{1}{8}f_1^s D_1^a +
	\frac{1}{8}f_1^a D_1^s +  \frac{3}{8}f_1^a D_1^a; 
\end{equation}}
\item{$ e\,n\,\rightarrow e'\,p\,X$
\begin{equation} 
\frac{4}{9} f_1^{n\rightarrow u} D_1^{u \rightarrow p} + 
\frac{1}{9} f_1^{n\rightarrow d} D_1^{d \rightarrow p} =
	\frac{1}{6}f_1^s D_1^a +\frac{2}{3}f_1^a D_1^s +  \frac{5}{18}f_1^a D_1^a. 
\end{equation}}
\end{itemize}
Analogous formulae apply to other combinations of distribution and
fragmentation functions.

In the following section we will concentrate only on the first process.
However, using appropriate coefficients $c_{ij}$ and appropriate hadron masses
one can carry out the calculations for any
baryon-to-baryon process. 

%%%%%%%%%%%%%%%%%%%%%%%%%%%%%%%%%%%%%%%%%%%%%%%%%%%%%%%%%%%%%%%%%%%%%%%%%%%%%%
\subsection{Numerical results for structure functions in $ e\,p\,\rightarrow
e'\,\Lambda\,X$}

In this section we present numerical results for structure functions of the
process $ e\,p\,\rightarrow e'\,\Lambda\,X$ obtained using the distribution
and fragmentation functions from the spectator model. We concentrate on
$\Lambda$ production, since it is possible to determine its polarization from
the kinematics of its decay.
We will consider separately different experimental situations with or without
polarization of the target and of the produced $\Lambda$. In this way the
various terms in Eq.~\ref{e:wt} and Eq.~\ref{e:wtt'} can be accessed. 

%%%%%%%%%%%%%%%%%%%%%%%%%%%%%%%%%%%%%%%%%%%
{\em 1 --    Unpolarized proton target and unpolarized produced $\Lambda$.}
We performed the calculation of the structure function
  $W_{T}^{\,\mbox{\tiny [UU]}}$ of Eq.~(\ref{e:wtuu}) 
using $f_1$ from Eq.~(\ref{e:f1}) and 
$D_1$ from Eq.~(\ref{e:d1}), with the coefficients $c_{ij}$ as 
specified in Eq.~(\ref{e:protlam}). 
After integrating over $k_x$ and $k_y$ by using the
$\delta$-function we obtain
\begin{eqnarray}  W_T^{\,\mbox{\tiny [UU]}}(\xbj,z,P_{h\perp}) &=&
	\sum_{i,j=s,a} c_{ij}\,
	{\textstyle \frac{ n_i^2 N_j^2}{2 M\, (2\pi)^6}}
	\;(1-\xbj)^3 \left(\frac{1-z}{z}\right)^3 
	\int \de p_x \de p_y \nn \\
    &&	\qquad\mbox{}\times
	\frac{(m + \xbj M)^2 +p_x^2 +p_y^2}
		{\left[p_x^2+p_y^2 + l_i^2(x) \right]^4}\;\;
	\frac{\left(m + \frac{1}{z}M_h\right)^2+
		\left(p_x-\frac{|P_{h\perp}|}{z}\right)^2 +p_y^2 }
	{\left[\left(p_x-\frac{|P_{h\perp}|}{z}\right)^2
			+p_y^2 + L_j^2(z) \right]^4}.
\end{eqnarray} 

The remaining integration has been carried out making use of an
adaptive multi-dimensional integration method. 
As we remarked before, integrating out $P_{h\perp}$ leads to a deconvolution
and, consequently, the transverse momentum integration can be performed
separately for the distribution and fragmentation functions as shown in 
Eq.~(\ref{e:wt2}). With the form of the functions given by the spectator model
the integrations can be carried out analytically~\cite{rod}. Comparison of
 those
analytical results with the outcome of a numerical integration has been used
as a check of consistency.

\begin{figure}
        \centering
        \epsfig{figure=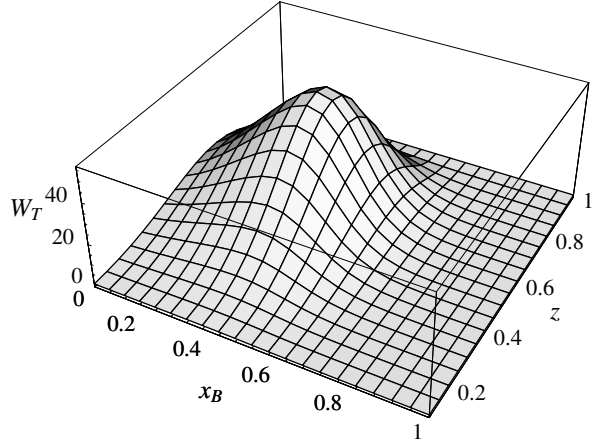,width=10cm}
        \caption{Dependence of the structure function 
		$W_T^{\,\mbox{\tiny[UU]}}$ on $\xbj$ and $z$ 
		at $P_{h\perp}=0$.}         
	\label{f:fdppxz}
        \end{figure}

\begin{figure}
	\centering
        \begin{tabular}{ccc}
	\epsfig{figure=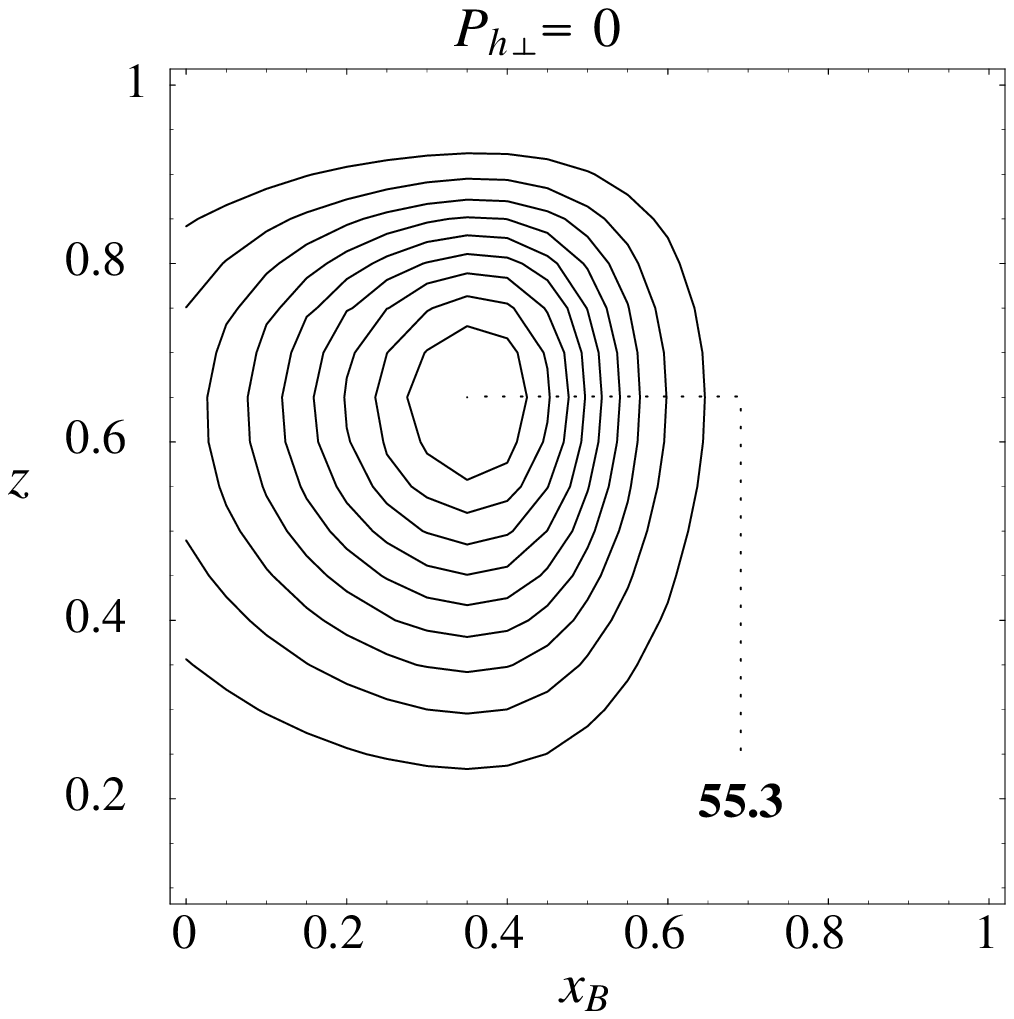,width=6cm}& 
			\epsfig{figure=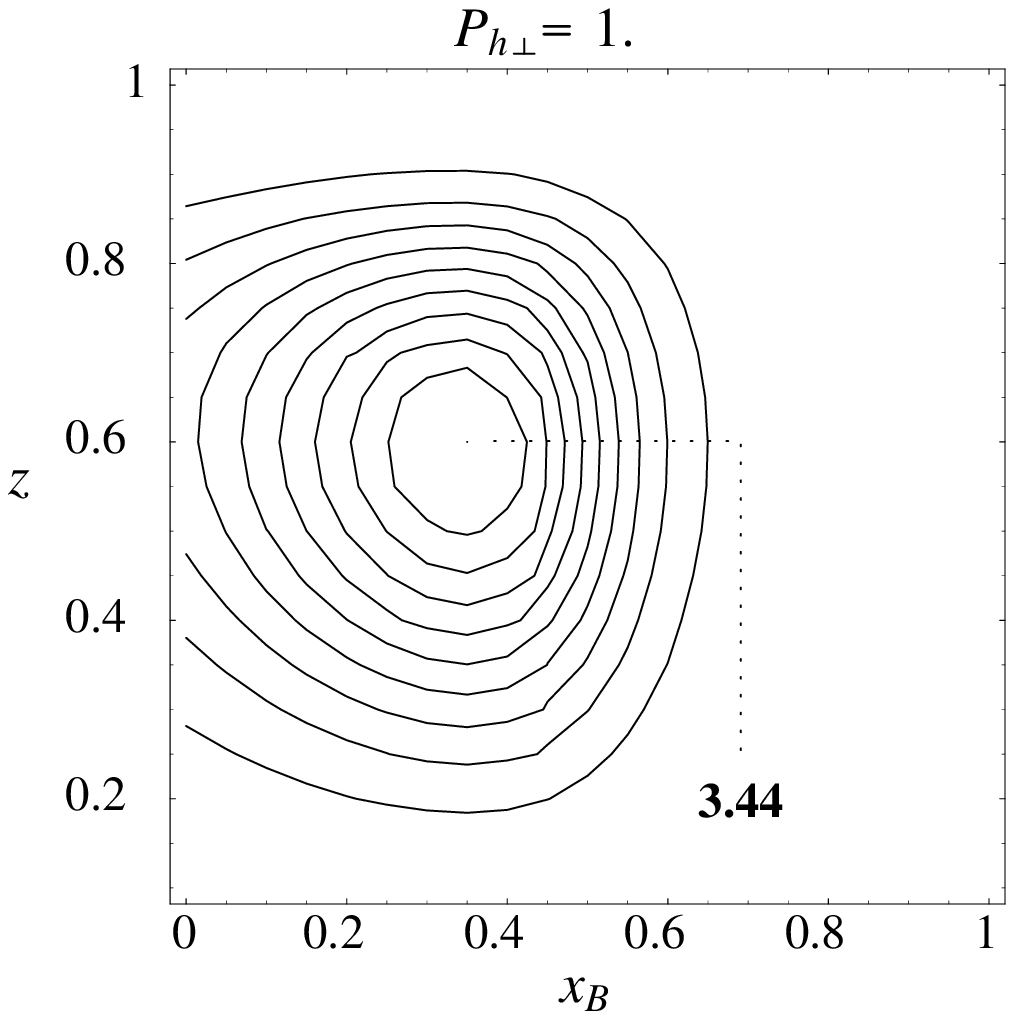,width=6cm}&
	\epsfig{figure=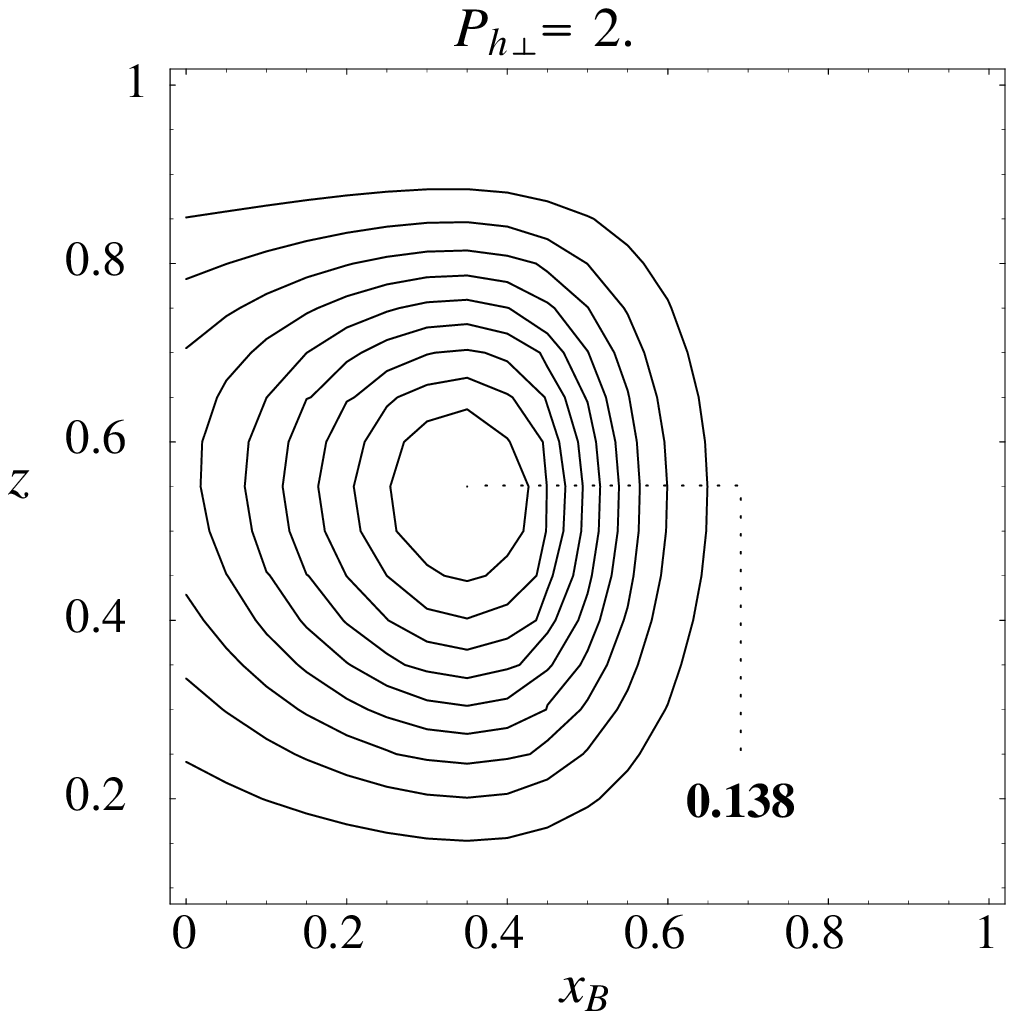,width=6cm}\\	
	\end{tabular}
        \caption{Contour-plot of the structure function 
	$W_T^{\,\mbox{\tiny [UU]}}$ for different
	values of $P_{h \perp}$ (GeV). Numbers inside the plot area denote the
	height of maxima. Spacing between isometric lines
	corresponds to 10~\% of maximum height.}
        \label{f:fdppxz1}
        \end{figure}

The results are displayed in the plots. Fig.~\ref{f:fdppxz} shows the 
structure function $W_T^{\,\mbox{\tiny [UU]}}$ at 
$P_{h\perp}=0$. Fig.~\ref{f:fdppxz1} shows the
contour-plot of the same function at three different values of
$P_{h\perp}$. An interesting feature is that when $P_{h\perp}$ increases, the 
position of the peak slowly moves to lower values of $z$  while there is no
change in the $x$ position of the peak. In other words, a hadron produced 
with a higher transverse momentum is more likely to carry a lower fraction of
the longitudinal momentum of the original quark. We remark that this behavior
is due to the kinematical conditions imposed by momentun conservation.

\begin{figure}
        \centering
        \begin{tabular}{cc}
	\epsfig{figure=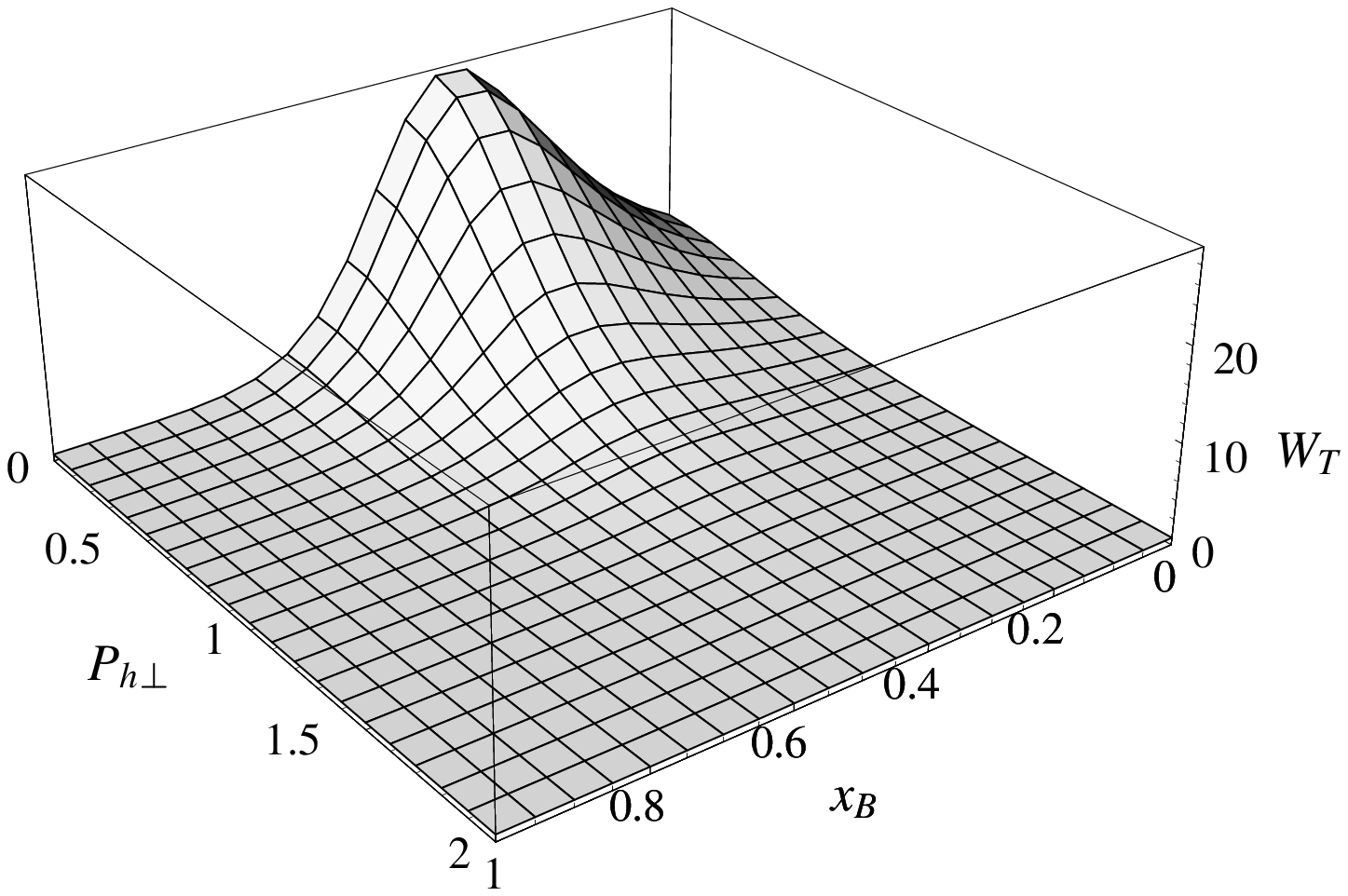,width=7cm}&
			\epsfig{figure=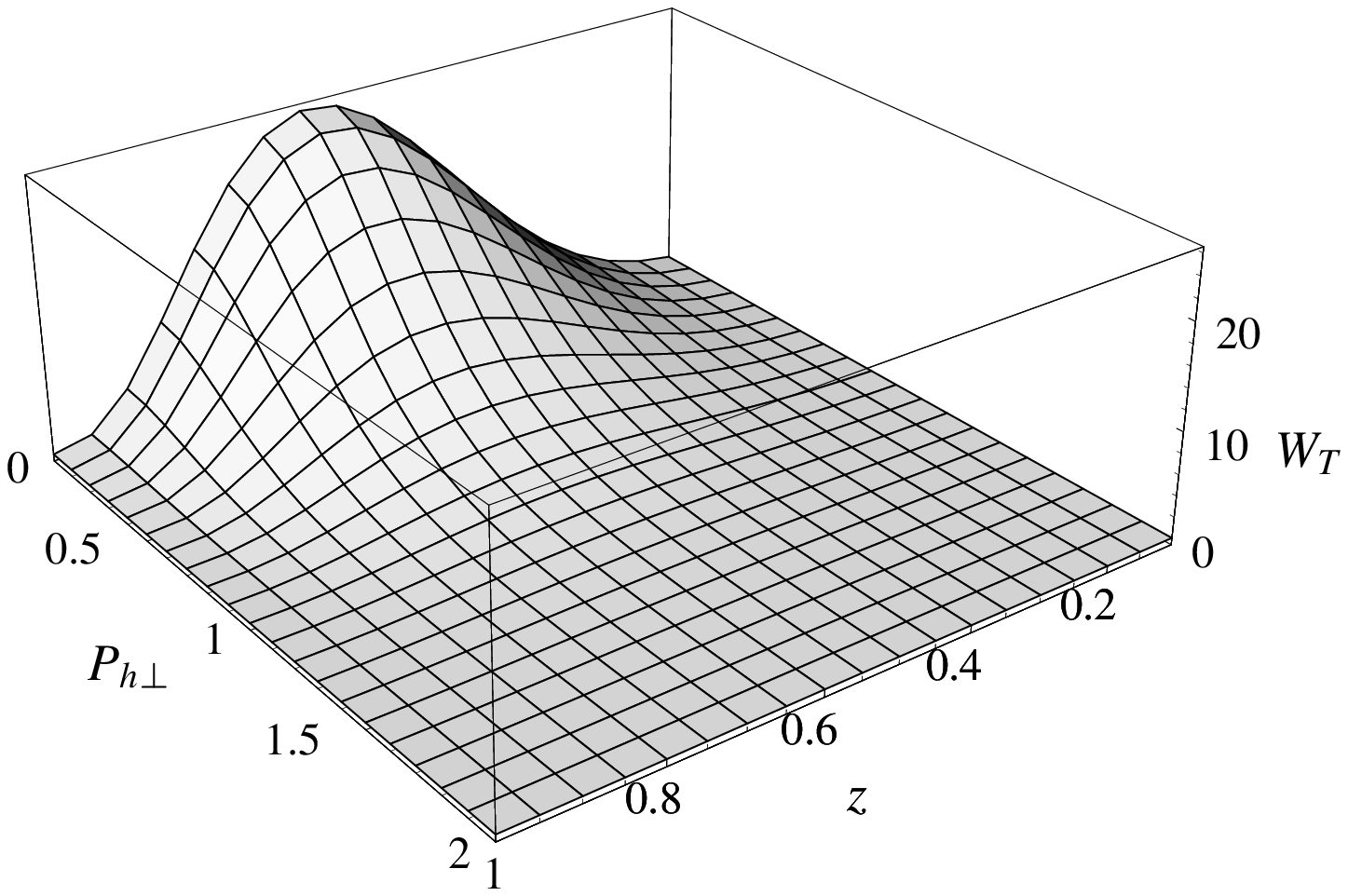,width=7cm}\\
	{\footnotesize (a)} &
			{\footnotesize (b)}\\ \\ 
	\end{tabular}
        \caption{Dependence of the structure function 
		$W_T^{\,\mbox{\tiny [UU]}}$ on the outgoing
		hadron transverse momentum, $P_{h \perp}$ (measured in GeV)
		and: (a) the fractional
		momentum $\xbj$, (b) the fractional momentum $z$. }
        \label{f:fdppx}
        \end{figure}

In Fig.~\ref{f:fdppx} we show the results obtained by
integrating the structure function over $z$ or $\xbj$, respectively.

\begin{figure}
        \centering
        \epsfig{figure=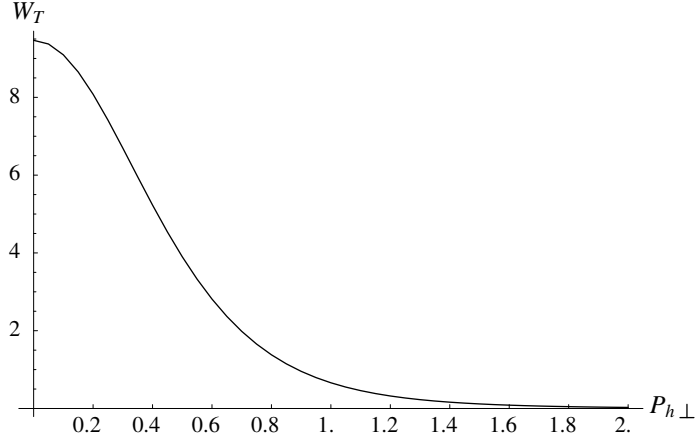,width=10cm}
        \caption{Dependence of the structure function 
		$W_T^{\,\mbox{\tiny [UU]}}$ on the outgoing
		hadron transverse momentum, $P_{h \perp}$ (GeV).}
        \label{f:fdpp}
        \end{figure}

Finally, we can integrate over both $z$ and $\xbj$ at the same time, thereby
obtaining the dependence of the structure function $W_T^{\,\mbox{\tiny [UU]}}$ 
on $P_{h \perp}$
alone (Fig.~\ref{f:fdpp}). This dependence is connected with 
the distribution of transverse momentum
of quarks inside the hadron. In principle, comparison with experimental data 
could be used to discriminate
between different assumptions on this distribution.

%%%%%%%%%%%%%%%%%%%%%%%%%%%%%%%%%%%%%%%%%
{\em 2 --   Longitudinally polarized proton target and unpolarized produced 
$\Lambda$.}
In this case, we need 
to calculate the structure function ${W'}_{TT}^{\,\mbox{\tiny [LU]}}$ of 
Eq.~(\ref{e:wtt'lu}). Using  Eq.~(\ref{e:g1l})
and Eq.~(\ref{e:d1}) and integrating over $k_x$ and $k_y$ using the
$\delta-$functions, we obtain the formula 
\begin{eqnarray} 
{W'}_{TT}^{\,\mbox{\tiny [LU]}}(\xbj,z,P_{h\perp}) 
    &=&\sum_{i,j=s,a} c_{ij}\,a_i\,
	 {\textstyle \frac{ n_i^2 N_j^2}{2 M\, (2\pi)^6}}
	\;(1-\xbj)^3 \left(\frac{1-z}{z}\right)^3 
	\int \de p_x \de p_y \nn \\
    &&	\qquad\mbox{}\times
	\frac{(m + \xbj M)^2 -p_x^2 -p_y^2}
		{\left[p_x^2+p_y^2 + l_i^2(x) \right]^4}\;\;
	\frac{\left(m + \frac{1}{z}M_h\right)^2 +
		\left(p_x-\frac{|P_{h\perp}|}{z}\right)^2 +p_y^2}
	{\left[\left(p_x-\frac{|P_{h\perp}|}{z}\right)^2
			+p_y^2 + L_j^2(z) \right]^4}.
\end{eqnarray} 

\begin{figure}
        \centering
        \epsfig{figure=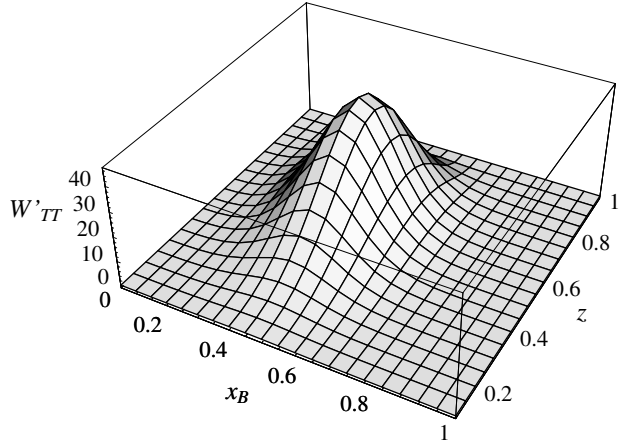,width=10cm}
        \caption{Dependence of the structure function 
		${W'}_{TT}^{\,\mbox{\tiny [LU]}}$ on $\xbj$ and $z$ 
		at $P_{h\perp}=0$.}
        \label{f:gldppxz}
        \end{figure}

\begin{figure}
	\centering
        \begin{tabular}{ccc}
	\epsfig{figure=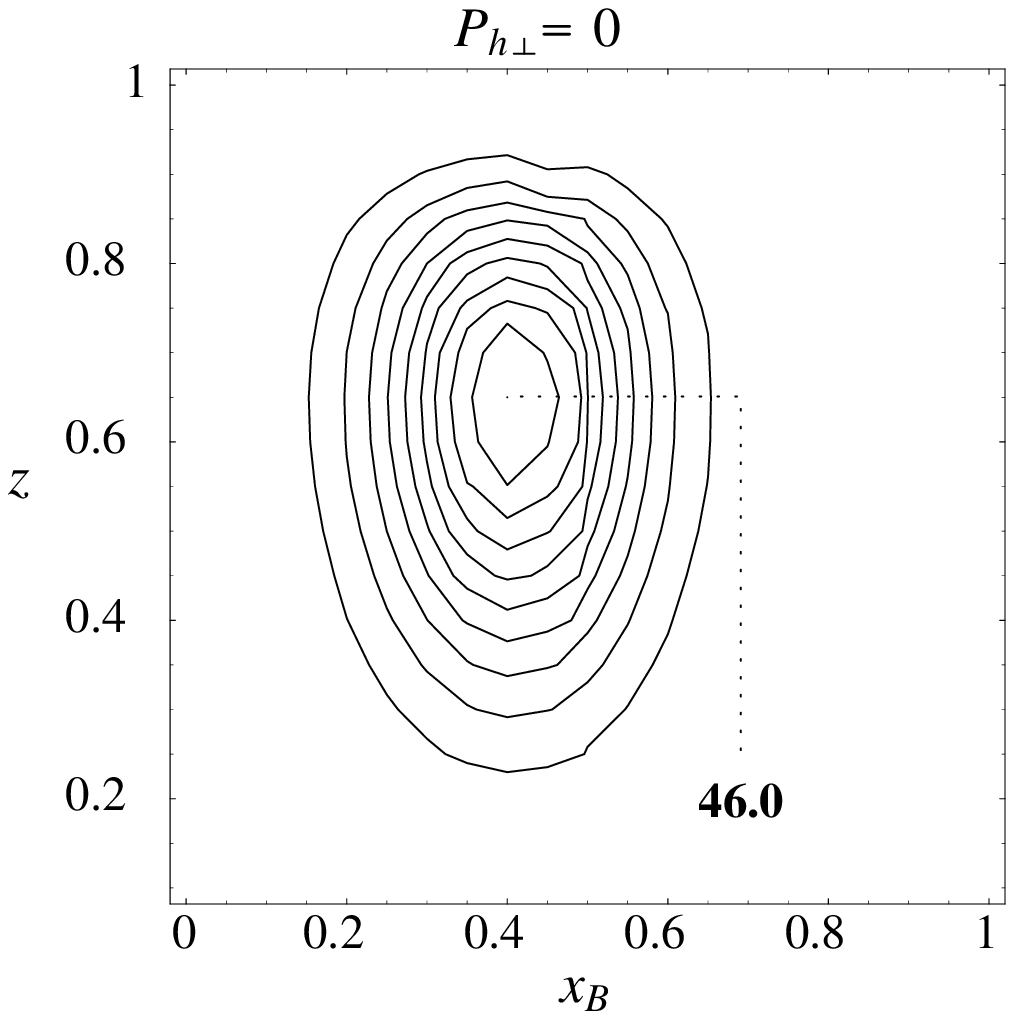,width=6cm}& 
			\epsfig{figure=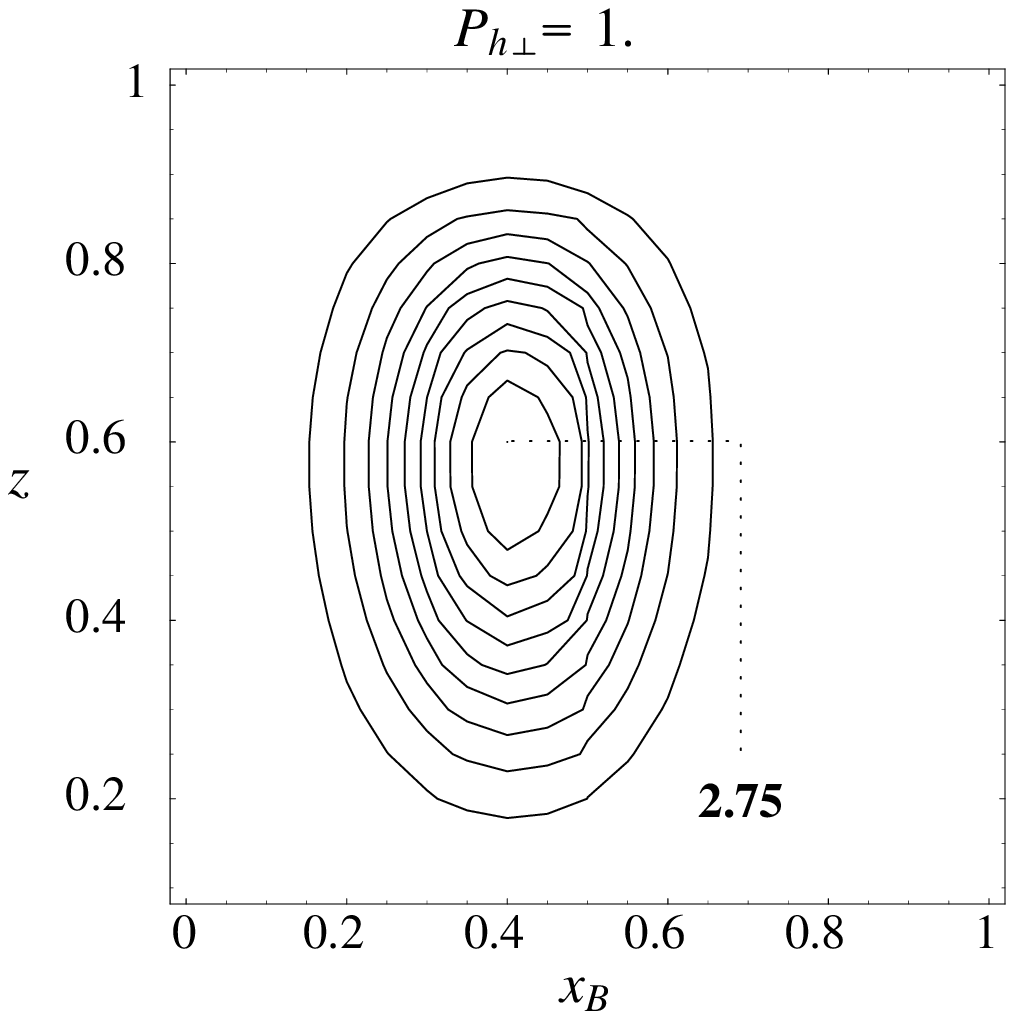,width=6cm}&
	\epsfig{figure=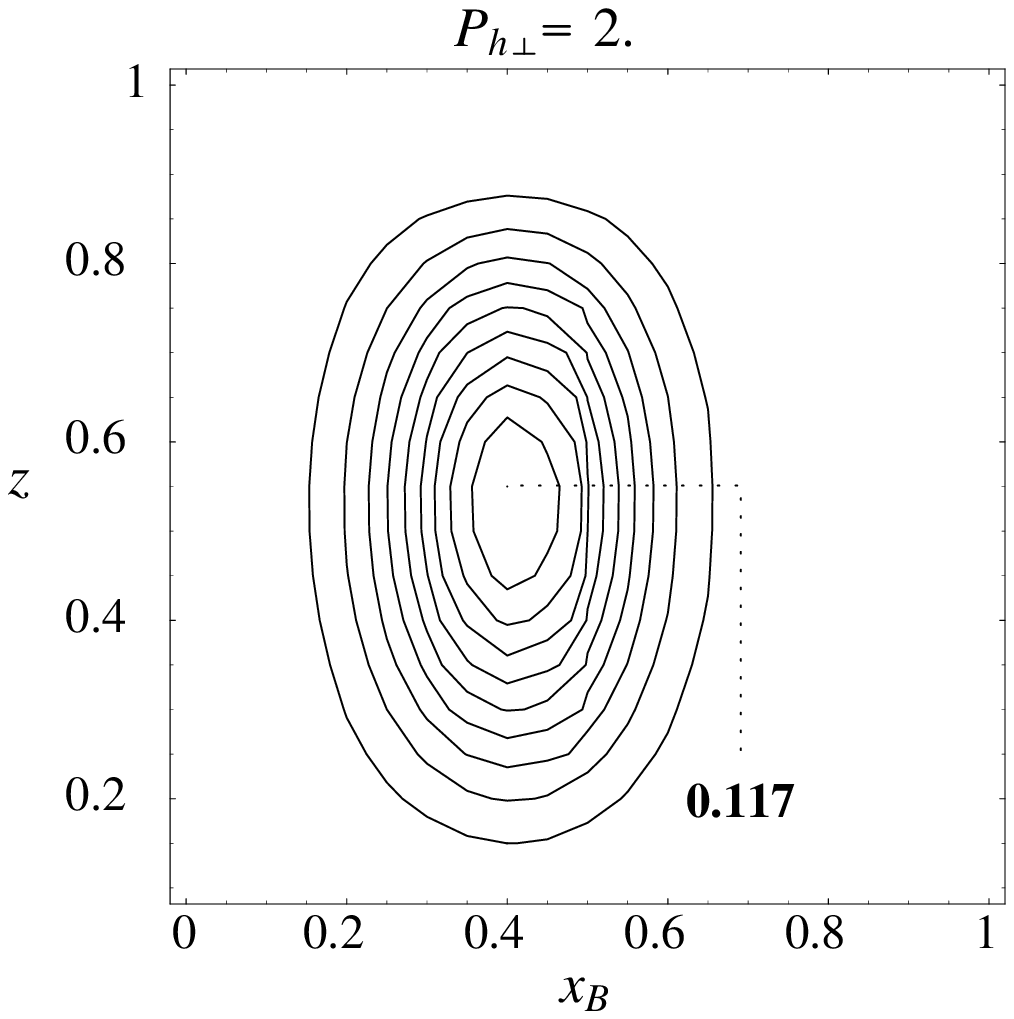,width=6cm}\\	
	\end{tabular}
        \caption{Contour-plot of the structure function 
	${W'}_{TT}^{\,\mbox{\tiny [LU]}}$ for different
	values of $P_{h \perp}$ (GeV). Numbers inside the plot area denote the
	height of maxima. Spacing between isometric lines
	corresponds to 10~\% of maximum height.}
        \label{f:gldppxz1}
        \end{figure}

Fig.~\ref{f:gldppxz} shows the 
structure function ${W'}_{TT}^{\,\mbox{\tiny [LU]}}$ at $P_{h\perp}=0$. 
Fig.~\ref{f:gldppxz1} shows the
contour-plot of the same function at three different values of
$P_{h\perp}$. As in the case of $W_T^{\,\mbox{\tiny [UU]}}$, for
increasing $P_{h\perp}$ the 
position of the peak moves to lower values of $z$.

Integration over $z$ or $\xbj$ produces a behavior similar to the one shown
in Fig.~\ref{f:fdppx}.

\begin{figure}
        \centering
        \epsfig{figure=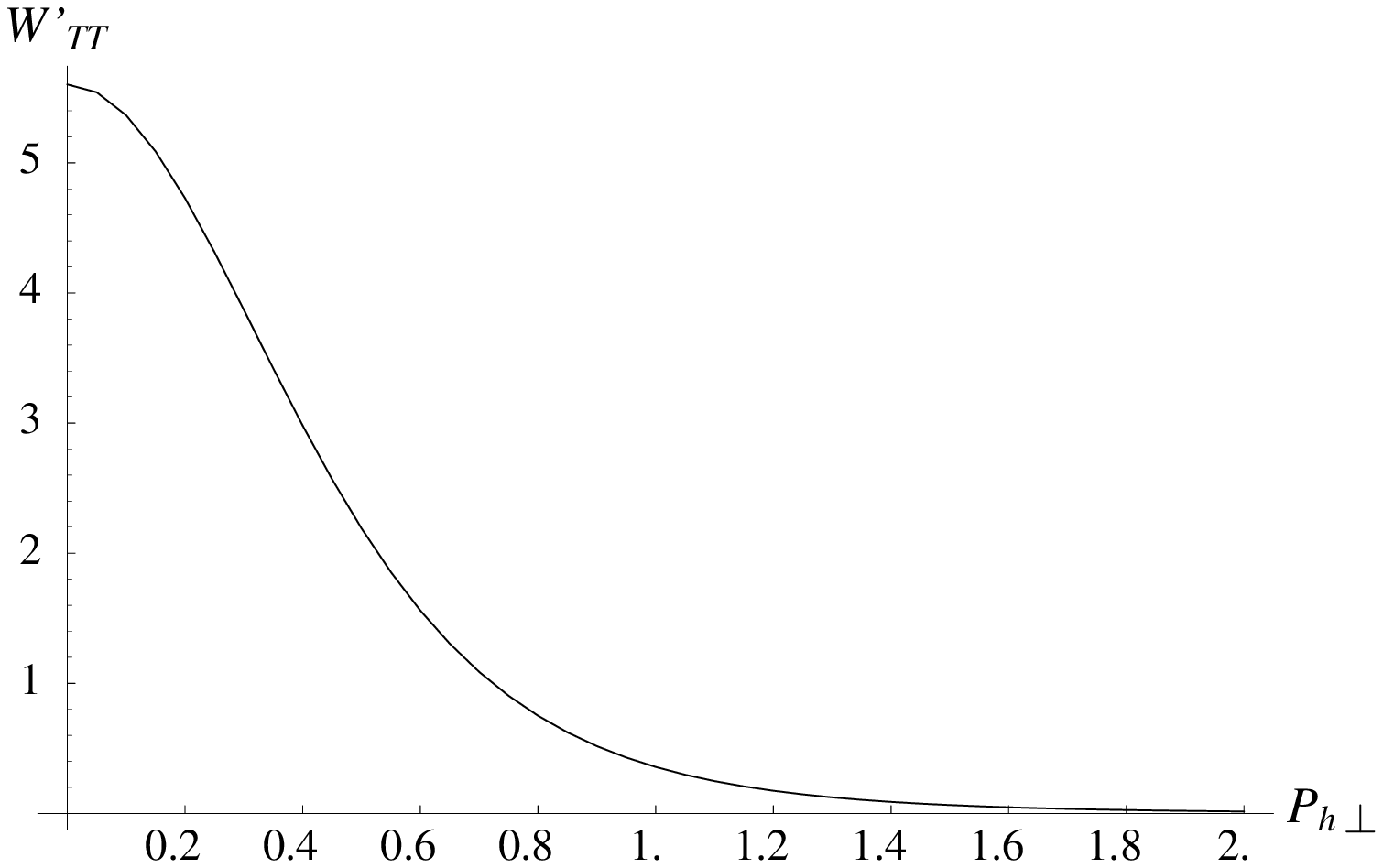,width=10cm}
        \caption{Dependence of the structure function 
		${W'}_{TT}^{\,\mbox{\tiny [LU]}}$ on the outgoing
		hadron transverse momentum, $P_{h \perp}$ (GeV).}
        \label{f:gldpp}
        \end{figure}

Integrating over both $z$ and $\xbj$ we
obtain the dependence of the structure function 
${W'}_{TT}^{\,\mbox{\tiny [LU]}}$ on $P_{h \perp}$ alone (Fig.~\ref{f:gldpp}).

%%%%%%%%%%%%%%%%%%%%%%%%%%%%%%%%%%%%%%%
{\em 3 --   Transversely polarized proton target and unpolarized produced 
$\Lambda$.}
Substituting the results coming from Eq.~(\ref{e:g1t}) and Eq.~(\ref{e:d1}) in
 Eq.~(\ref{e:wtt'tu}) and integrating over $k_x$ and $k_y$ 
we obtain
\begin{eqnarray} 
{W'}_{TT}^{\,\mbox{\tiny [TU]}}(\xbj,z,P_{h\perp}) 
    &=&\sum_{i,j=s,a} c_{ij}\,a_i\,
	 {\textstyle \frac{ n_i^2 N_j^2}{2 M\, (2\pi)^6}}
	\;(1-\xbj)^3 \left(\frac{1-z}{z}\right)^3 
	\int \de p_x \de p_y \nn \\
    &&	\qquad\mbox{}\times
	\frac{2\, p_x\,(m + \xbj M) }
		{\left[p_x^2+p_y^2 + l_i^2(x) \right]^4}\;\;
	\frac{\left(m + \frac{1}{z}M_h\right)^2+
		\left(p_x-\frac{|P_{h\perp}|}{z}\right)^2 +p_y^2 }
	{\left[\left(p_x-\frac{|P_{h\perp}|}{z}\right)^2
			+p_y^2 + L_j^2(z) \right]^4}.
\end{eqnarray}

\begin{figure}
        \centering
        \epsfig{figure=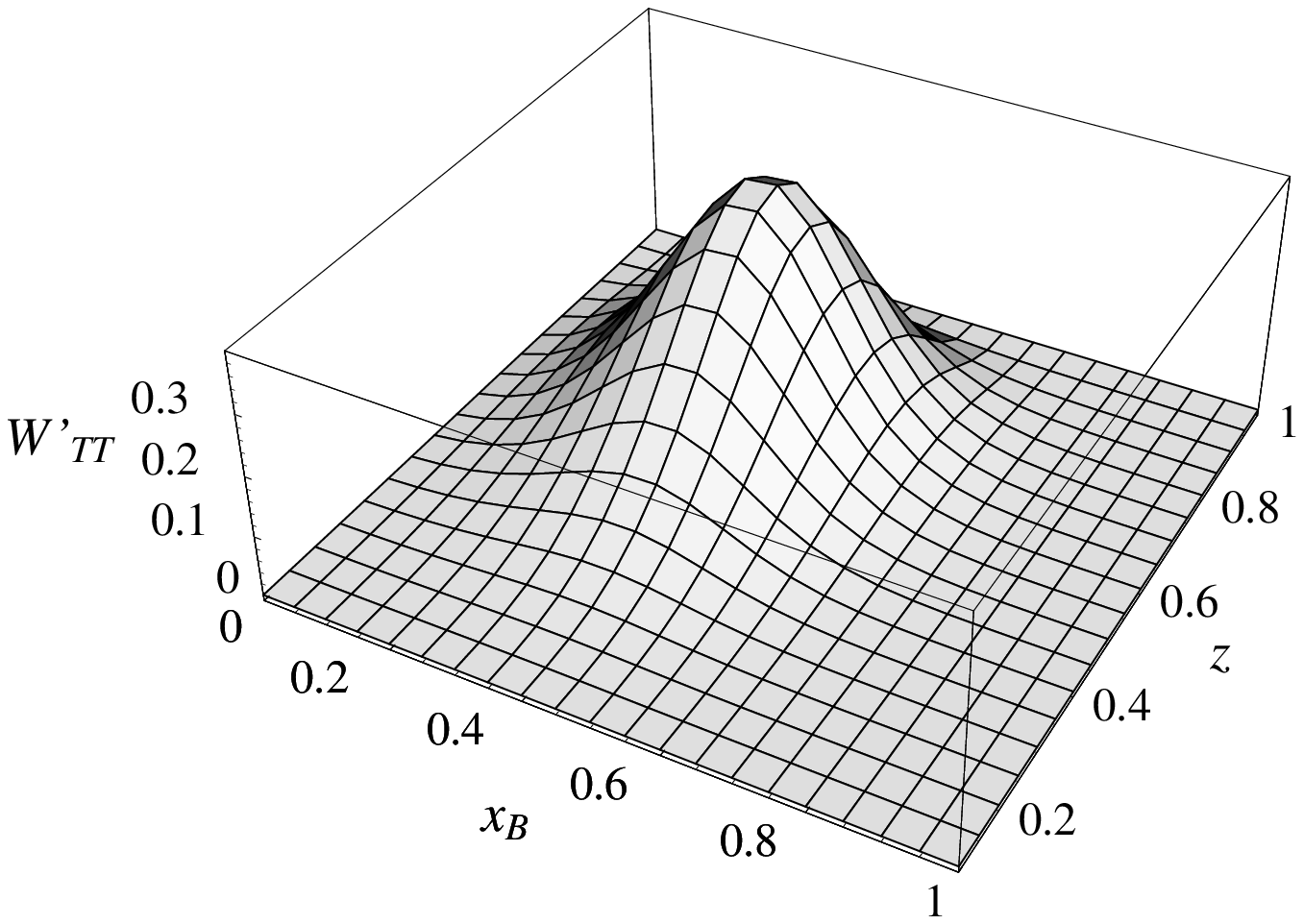,width=10cm}
        \caption{Dependence of the structure function 
		${W'}_{TT}^{\,\mbox{\tiny [TU]}}$ on $\xbj$ and 
		$z$ at $P_{h \perp}=0.4$ GeV. }
        \label{f:gtdppxz}
        \end{figure}

\begin{figure}
	\centering
        \begin{tabular}{ccc}
	\epsfig{figure=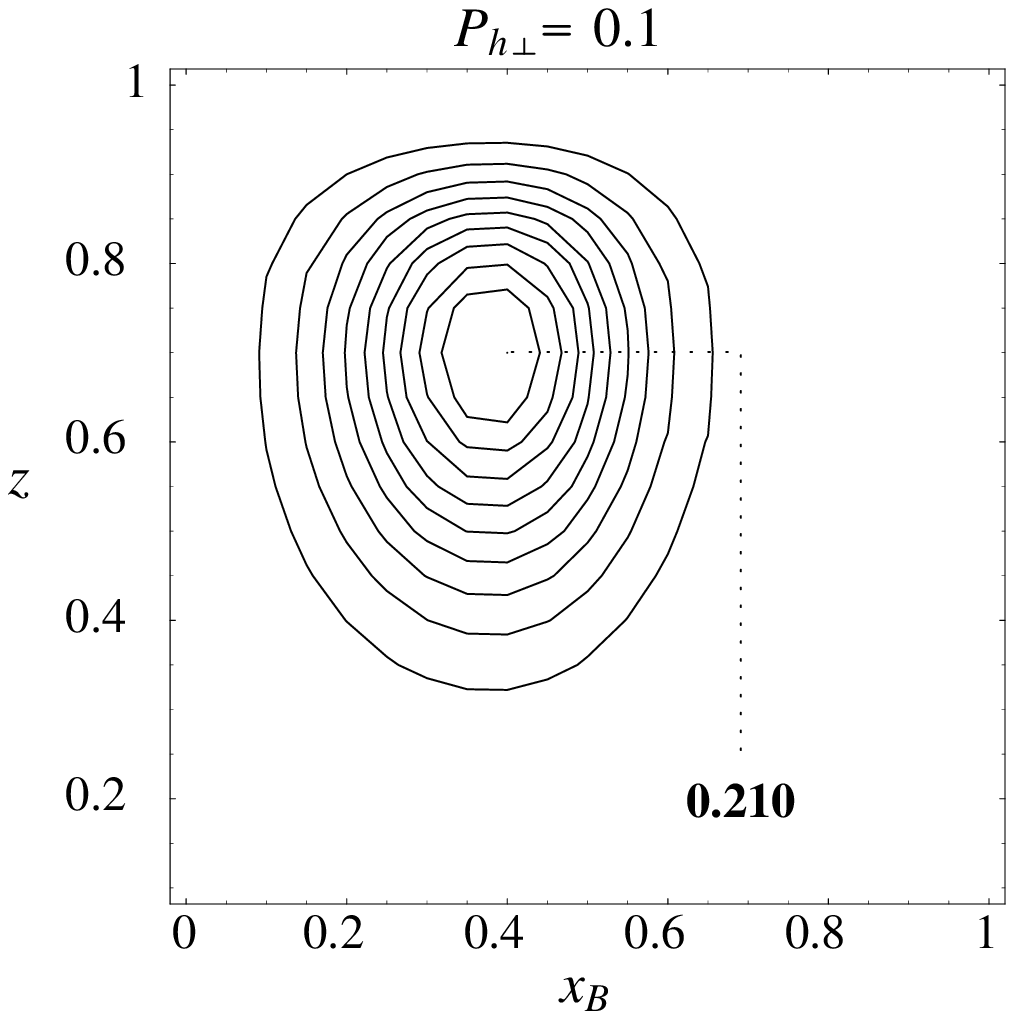,width=6cm}& 
			\epsfig{figure=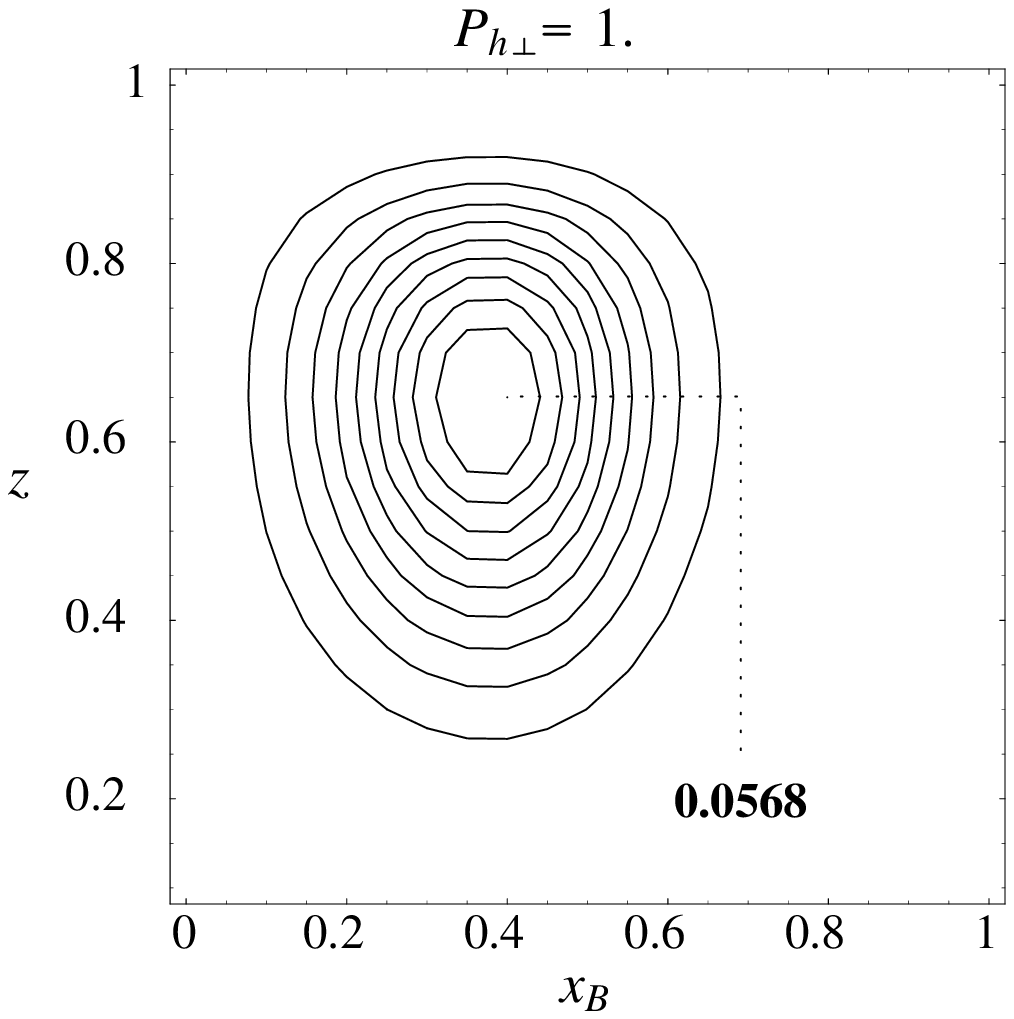,width=6cm}&
	\epsfig{figure=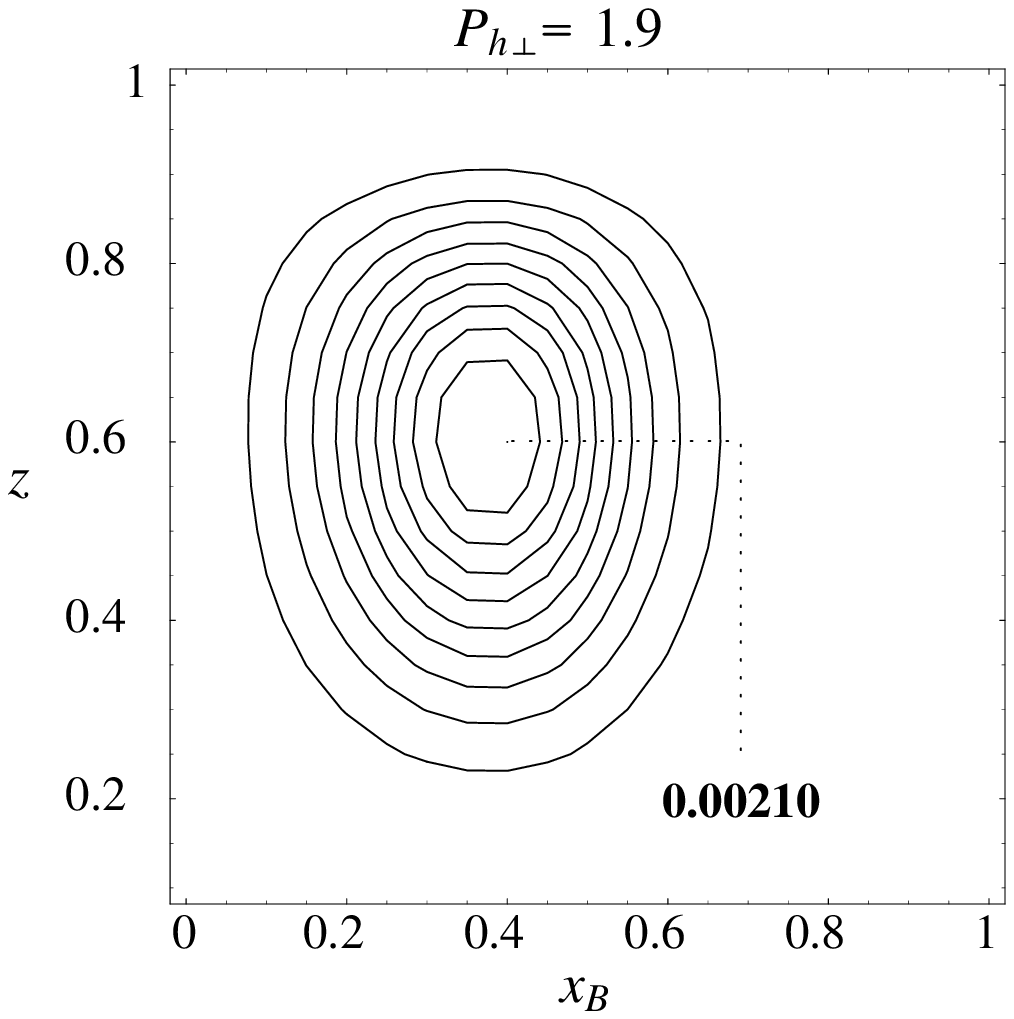,width=6cm}\\	
	\end{tabular}
        \caption{Contour-plot of the structure function 
	${W'}_{TT}^{\,\mbox{\tiny [TU]}}$ for different
	values of $P_{h \perp}$ (GeV). Numbers inside the plot area denote the
	height of maxima. Spacing between isometric lines
	corresponds to 10~\% of maximum height.}
        \label{f:gtdppxz1}
        \end{figure}

Fig.~\ref{f:gtdppxz} shows the 
structure function ${W'}_{TT}^{\,\mbox{\tiny [TU]}}$ at $P_{h\perp}=0.4$ GeV. 
Fig.~\ref{f:gtdppxz1} shows the 
contour-plot of the same function at three different values of
$P_{h\perp}$. Again, as $P_{h\perp}$ increases the 
position of the peak moves to lower values of $z$.

\begin{figure}
        \centering
        \epsfig{figure=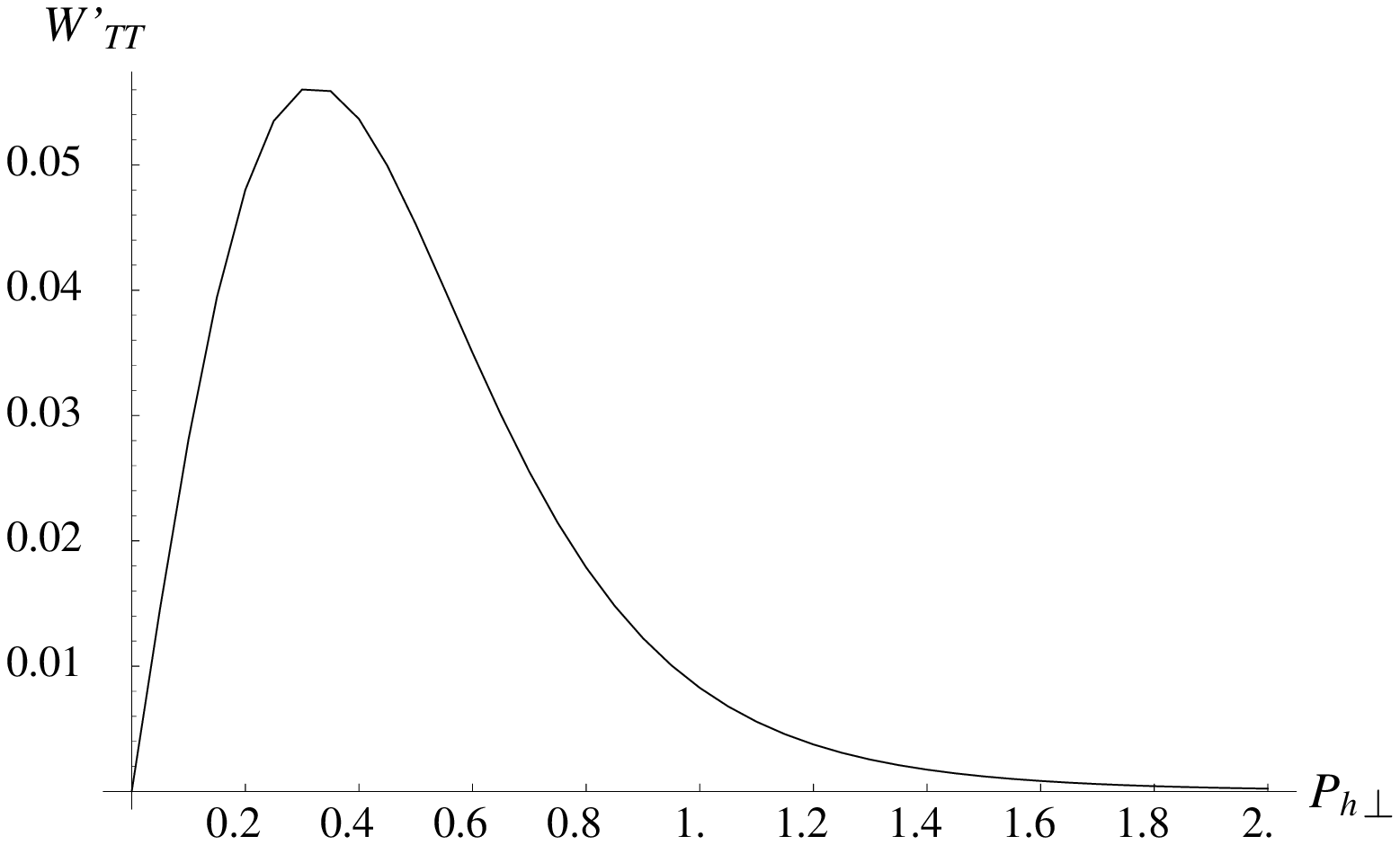,width=10cm}
        \caption{Dependence of the structure function 
		${W'}_{TT}^{\,\mbox{\tiny [TU]}}$ on the outgoing
		hadron transverse momentum, $P_{h \perp}$ (GeV).}
        \label{f:gtdpp}
        \end{figure}

Integrating over both $z$ and $\xbj$ we
obtain the dependence of the structure function 
${W'}_{TT}^{\,\mbox{\tiny [TU]}}$ on $P_{h \perp}$ alone (Fig.~\ref{f:gtdpp}).

%%%%%%%%%%%%%%%%%%%%%%%%%%%%%%%%%%%%%%%%%%%%%
{\em 4 --   Unpolarized proton target and longitudinally polarized produced 
$\Lambda$.}
Substituting the results coming from Eq.~(\ref{e:f1}) and Eq.~(\ref{e:g1l}) in
 Eq.~(\ref{e:wtt'ul}) and integrating over $k_x$ and $k_y$ 
we obtain
\begin{eqnarray} 
{W'}_{TT}^{\,\mbox{\tiny [UL]}}(\xbj,z,P_{h\perp}) 
    &=&	\sum_{i,j=s,a} c_{ij}\,a_j\,
	 {\textstyle \frac{ n_i^2 N_j^2}{2 M\, (2\pi)^6}}
	\;(1-\xbj)^3 \left(\frac{1-z}{z}\right)^3 
	\int \de p_x \de p_y \nn \\
    &&	\qquad\mbox{}\times
	\frac{(m + \xbj M)^2 + p_x^2 +p_y^2 }
		{\left[p_x^2+p_y^2 + l_i^2(x) \right]^4}\;\;
	\frac{\left(m + \frac{1}{z}M_h\right)^2 
			-\left(p_x-\frac{|P_{h\perp}|}{z}\right)^2 -p_y^2 }
	{\left[\left(p_x-\frac{|P_{h\perp}|}{z}\right)^2
			+p_y^2 + L_j^2(z) \right]^4}. 
\end{eqnarray}

\begin{figure}
        \centering
        \begin{tabular}{cc}
	\epsfig{figure=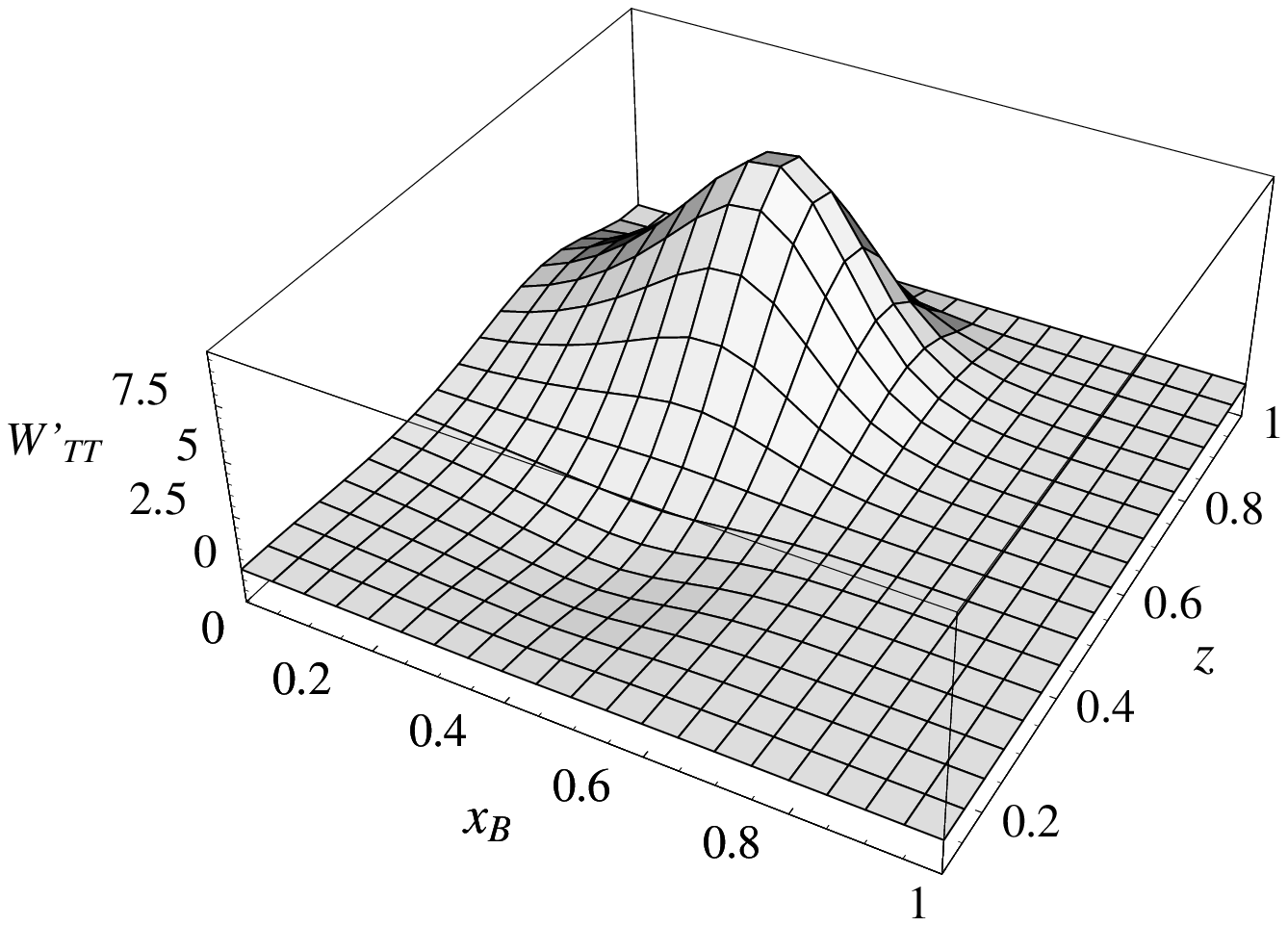,width=8cm}&
			\epsfig{figure=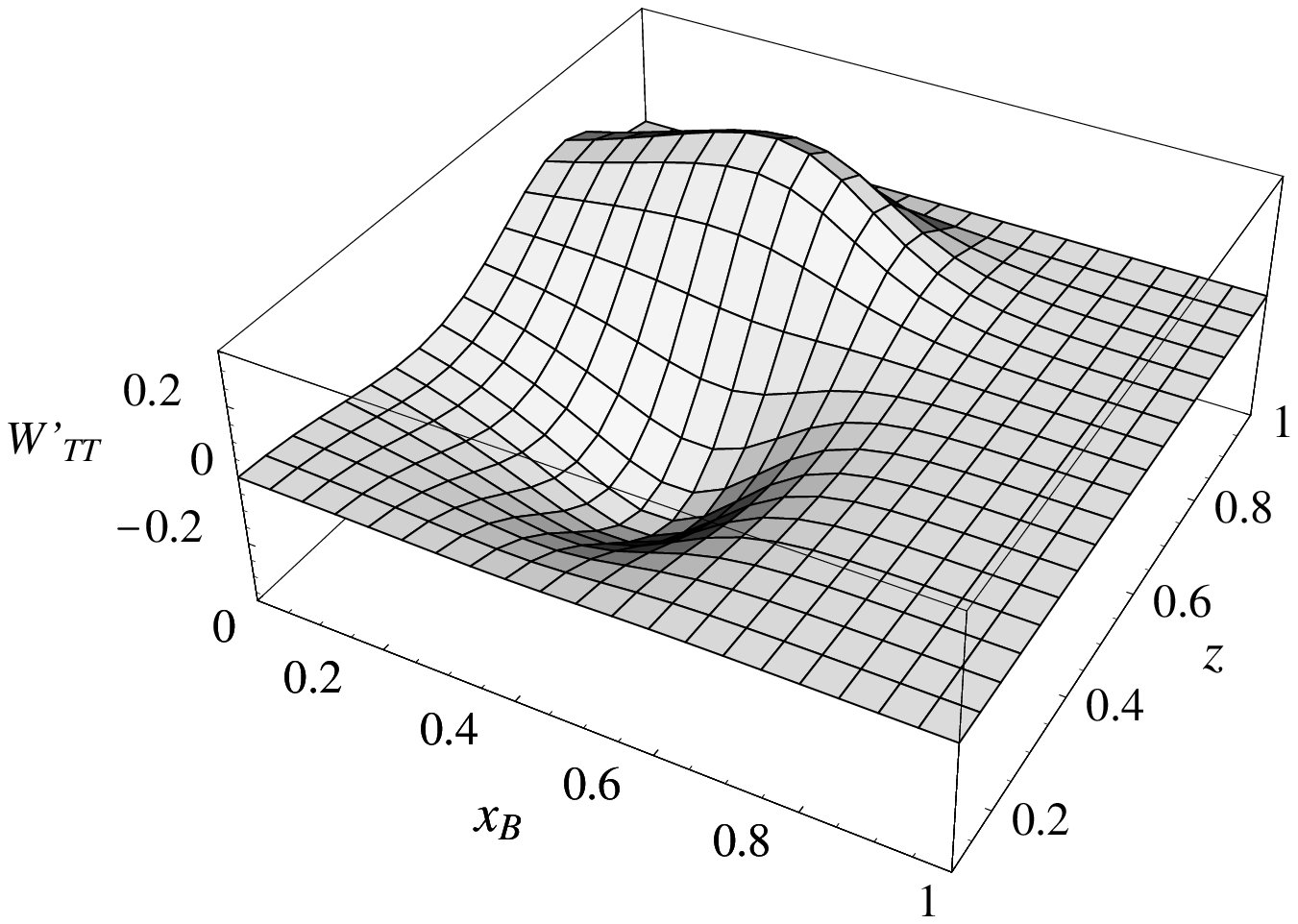,width=8cm}\\
	{\footnotesize $P_{h \perp}=0$} &
			{\footnotesize $P_{h \perp}=0.7$ GeV}\\ \\ 
	\end{tabular}
	\caption{Dependence of the structure function 
		${W'}_{TT}^{\,\mbox{\tiny [UL]}}$ on $\xbj$ and 
		$z$ at two different values of $P_{h \perp}$. }
        \label{f:fglppxz}
        \end{figure}

Fig.~\ref{f:fglppxz} shows the 
structure function ${W'}_{TT}^{\,\mbox{\tiny [UL]}}$ at $P_{h\perp}=0$ and at 
$P_{h\perp}=0.5$ GeV. 
In this case, the
contributions containing the negative $a_a$ factor play a larger role than in
the previous cases. For
this reason, the significance of the contour plots is reduced and we preferred
to show 3-D plots of the structure function.   
%Fig.~\ref{f:fglppxz1} shows the
%contour-plot of the same function at three different values of
%$P_{h\perp}$. As in the previous case, as $P_{h\perp}$ increases the 
%position of the peak moves to lower values of $z$.

\begin{figure}
        \centering
        \epsfig{figure=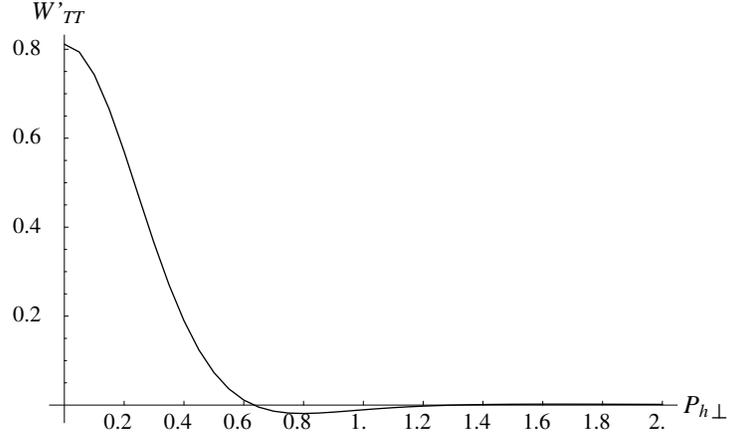,width=10cm}
        \caption{Dependence of the structure function 
		${W'}_{TT}^{\,\mbox{\tiny [UL]}}$ on the outgoing
		hadron transverse momentum, $P_{h \perp}$ (GeV).}
        \label{f:fglpp}
        \end{figure}

Integrating over both $z$ and $\xbj$ we
obtain the dependence of the structure function 
${W'}_{TT}^{\,\mbox{\tiny [UL]}}$ on $P_{h \perp}$ alone (Fig.~\ref{f:fglpp}).

%%%%%%%%%%%%%%%%%%%%%%%%%%%%%%%%%%%%%%%%%%%%%%%%%%%%%%%%%5
{\em 5 --   Unpolarized proton target and transversely polarized produced 
$\Lambda$.}
Substituting the results coming from Eq.~(\ref{e:f1}) and Eq.~(\ref{e:g1t}) in
the fourth term of Eq.~(\ref{e:wtt'}) and integrating over $k_x$ and $k_y$ 
we obtain
\begin{eqnarray} 
{W'}_{TT}^{\,\mbox{\tiny [UT]}}(\xbj,z,P_{h\perp}) 
    &=&	\sum_{i,j=s,a} c_{ij}
	 {\textstyle \frac{ n_i^2 N_j^2}{2 M\, (2\pi)^6}}
	\;(1-\xbj)^3 \left(\frac{1-z}{z}\right)^3 
	\int \de p_x \de p_y \nn \\
    &&	\qquad\mbox{}\times
	\frac{ p_x^2 +p_y^2 +(m + \xbj M)^2 }
		{\left[p_x^2+p_y^2 + l_i^2(x) \right]^4}\;\;
	\frac{2\,\left(p_x-\frac{|P_{h\perp}|}{z}\right)
					\left(m + \frac{1}{z}M_h\right) }
	{\left[\left(p_x-\frac{|P_{h\perp}|}{z}\right)^2
			+p_y^2 + L_j^2(z) \right]^4}. 
\end{eqnarray}

\begin{figure}
        \centering
        \begin{tabular}{cc}
	\epsfig{figure=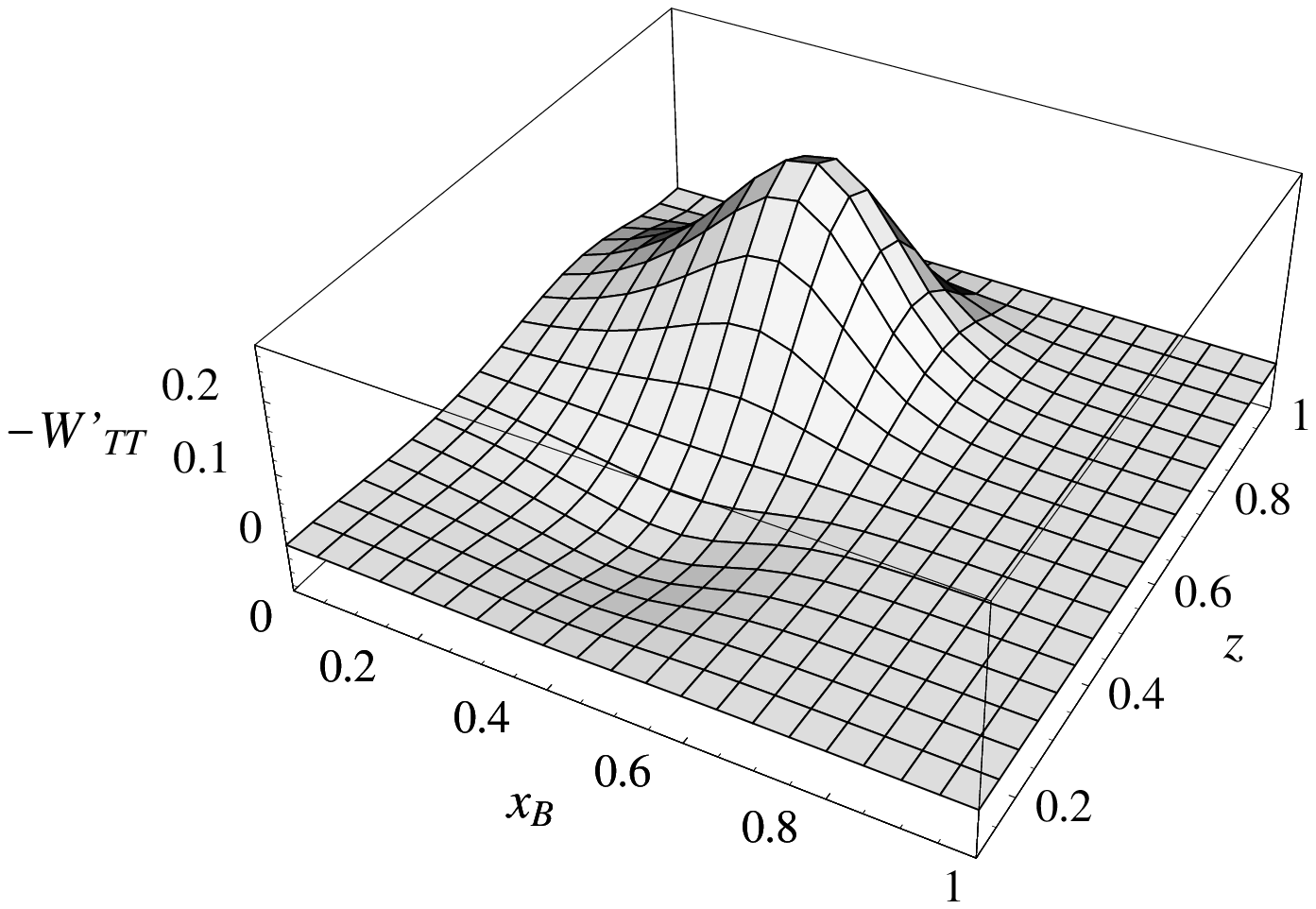,width=8cm}&
			\epsfig{figure=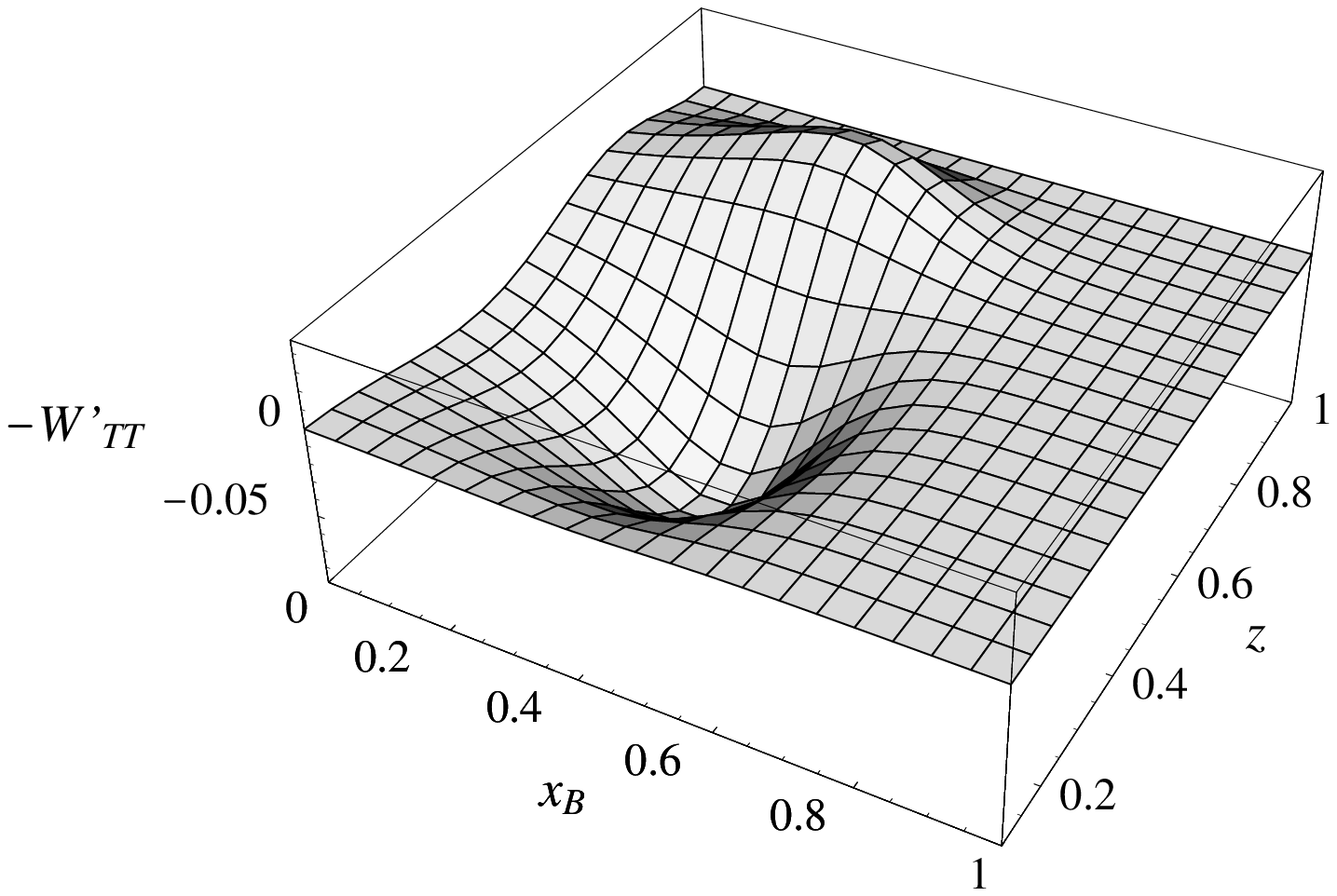,width=8cm}\\
	{\footnotesize $P_{h \perp}=0.4$ GeV} & 
			{\footnotesize $P_{h \perp}=0.7$ GeV}\\ \\ 
	\end{tabular}
	\caption{Dependence of the structure function 
		$-{W'}_{TT}^{\,\mbox{\tiny [UT]}}$ on $\xbj$ and 
		$z$ at two different values of $P_{h \perp}$. }
        \label{f:fgtppxz}
        \end{figure}

Fig.~\ref{f:fgtppxz} shows the 
structure function ${W'}_{TT}^{\,\mbox{\tiny [UT]}}$ 
(we plot $-{W'}_{TT}^{\,\mbox{\tiny [UT]}}$ for display convenience) 
at $P_{h\perp}=0.4$ GeV and $P_{h\perp}=0.8$ GeV. 
%Fig.~\ref{f:fgtppxz1} shows the
%contour-plot of the same function at three different values of
%$P_{h\perp}$. As in the previous case, as $P_{h\perp}$ increases the 
%position of the peak moves to lower values of $z$.

\begin{figure}
        \centering
        \epsfig{figure=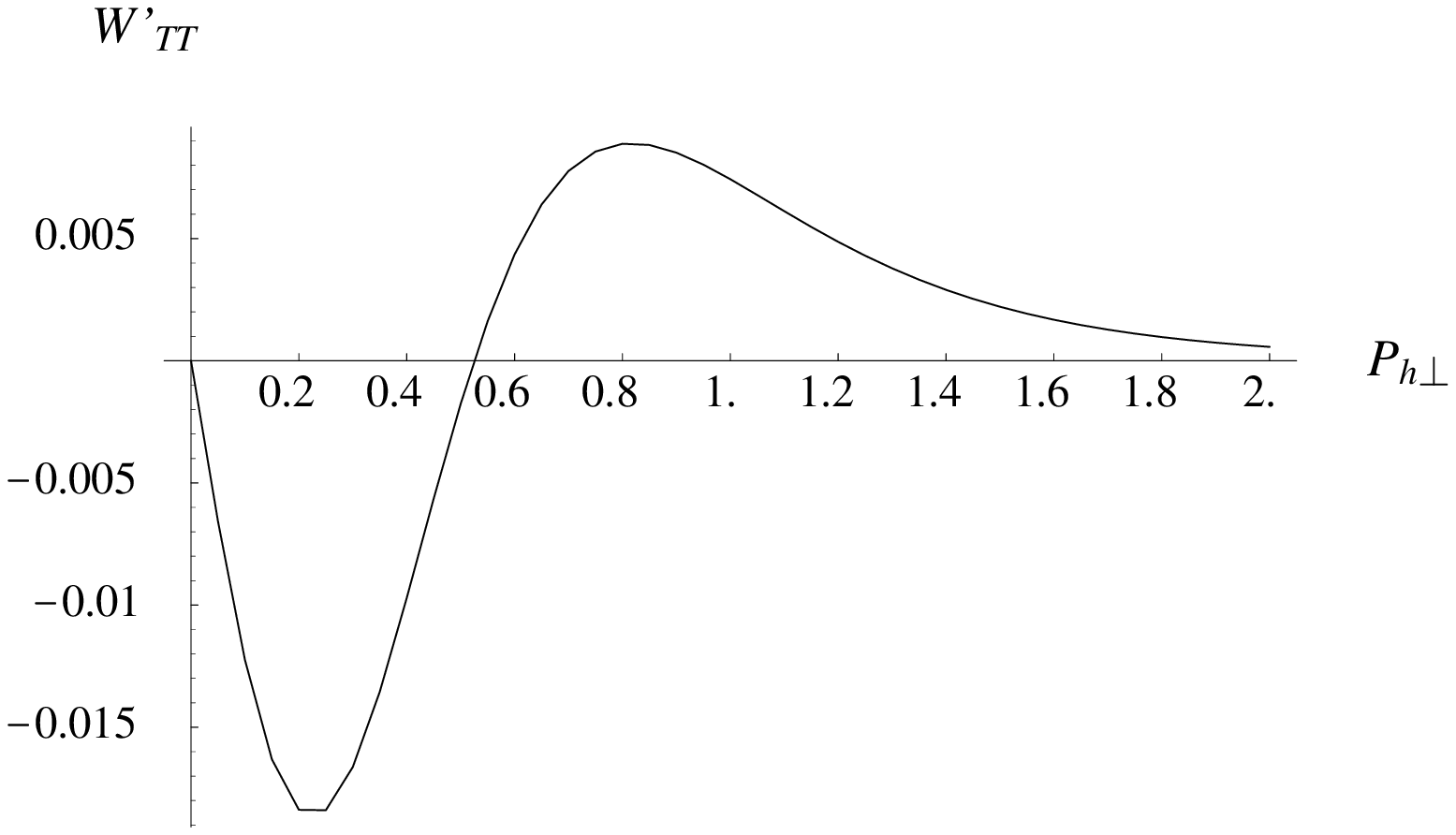,width=10cm}
        \caption{Dependence of the structure function 
		${W'}_{TT}^{\,\mbox{\tiny [UT]}}$ on the outgoing
		hadron transverse momentum, $P_{h \perp}$ (GeV).}
        \label{f:fgtpp}
        \end{figure}

Integrating over both $z$ and $\xbj$ we
obtain the dependence of the structure function 
${W'}_{TT}^{\,\mbox{\tiny [UT]}}$ on $P_{h \perp}$ alone (Fig.~\ref{f:fgtpp}).

%%%%%%%%%%%%%%%%%%%%%%%%%%%%%%%%%%%%%%%%%%%%%%%%%%%%%%%%%%%%%%%%%%%%%%%%%
\section{Conclusions}

In this paper the structure functions appearing in the cross section of
polarized one-particle inclusive deep-inelastic scattering have been expressed
in terms of distribution and fragmentation functions. Suitable kinematic
conditions allow to extract selected structure functions and thus to access
particular combinations of distribution and fragmentation functions.

We made use of a simple spectator model to calculate the
relevant structure functions involved in the cross section.
Our analysis can be applied to any process involving a baryonic
spin-$\frac{1}{2}$ target and an outgoing spin-$\frac{1}{2}$ baryon.
We chose to focus our attention on the process $e p \rightarrow e' \Lambda X$.
Using our results it is possible to estimate cross-sections for the cases when
proton and $\Lambda$ are both unpolarized or when one of the two is polarized.

An important feature of the analysis we presented is that we calculated the
dependence of the cross-sections on the transverse momentum of the outgoing
hadron, $P_{h\perp}$. The measurement of this variable gives access to two new
contributions to the structure functions. These contributions have never been
observed so far, because they vanish if the cross-section is integrated over
$P_{h\perp}$. 
Furthermore, the dependence of the cross-section on $P_{h\perp}$ indirectly
tests the distribution of partonic transverse momentum inside the hadron. This
distribution is largely unknown at the moment. 

The spectator model allows to study $P_{h\perp}$-dependent cross-sections
because it produces a well-defined, analytical form of transverse momentum
distributions.  Since the model qualitatively agrees with totally inclusive
measurements, we expect it to give good hints on the $P_{h\perp}$ dependence,
as well.

In summary, we are confident that the presented estimate can reproduce the
broad features of the structure functions observable in semi-inclusive deep
inelastic scattering.

%%%%%%%%%%%%%%%%%%%%%%%%%%%%%%%%%%%%%%%%%%%%%%%%%%%%%%%%%%%%%%%%%%%%%%%%%%
\acknowledgments{This work is supported by the Foundation for Fundamental
Research on Matter (FOM) and the Dutch Organization for Scientific Research 
(NWO) and was partly performed under contract ERB FMRX-CT96-0008 within the
frame of the Training and Mobility of Researchers Program of
the European Union.}

%%%%%%%%%%%%%%%%%%%%%%%%%%%%%%%%%%%%%%%%%%%%%%%%%%%%%%%%%%%%%%%%%%%%%%%%%%

\end{document}